\documentclass[%
 reprint,
 amsmath,amssymb, onecolumn
aps,
pre,
]{revtex4-2}

\usepackage{graphicx}
\usepackage{dcolumn}
\usepackage{bm}

\usepackage{xcolor}
\usepackage{ulem}

\begin{document}

\title{A diffuse-interface theory of active nematic interfaces: transport mechanisms and modal structure}

\author{Rodrigo C. V. Coelho$^{1,2}$}
\author{Mykola Tasinkevych$^{2,3,4}$}
\author{Margarida M. Telo da Gama$^{2,3,4}$}

\affiliation{$^1$Centro Brasileiro de Pesquisas Físicas, Rua Xavier Sigaud 150, 22290-180 Rio de Janeiro, Brazil.}
 \affiliation{$^2$Centro de Física Teórica e Computacional, Faculdade de Ciências, Universidade de Lisboa, 1749-016 Lisboa, Portugal.}
 \affiliation{$^3$Departamento de Física, Faculdade de Ciências, Universidade de Lisboa, P-1749-016 Lisboa, Portugal.}
\affiliation{$^4$International Institute for Sustainability with Knotted Chiral Meta Matter, Hiroshima University, Higashihiroshima 739-8511, Japan.}

\begin{abstract}
We develop a long-wavelength theory for the linear stability of a flat
interface between an active nematic and an isotropic fluid. Starting from
a diffuse-interface Cahn--Hilliard--Landau--de Gennes description coupled
to Brinkman-screened Stokes hydrodynamics, we project the linearized
dynamics onto a small set of interfacial degrees of freedom: the conserved
translation, or height, mode; a scalar profile distortion or amplitude mode;
and a transverse orientational mode associated with director rotations.
Eliminating the gapped scalar profile mode gives a reduced interfacial
operator coupling the conserved height mode to the transverse orientational
mode.

The main result is that activity generates, in the screened
diffuse-interface regime, a direct local contribution proportional to
\(q^2\) in the height sector. This term competes with the passive local
diffusive capillary relaxation, which enters at order \(q^4\), and defines
a local active interfacial channel controlled by the internal structure of
the diffuse interface. This mechanism is distinct from the non-analytic
\(|q|\) and \(|q|q^2\) terms characteristic of weakly screened
Hele--Shaw/Saffman--Taylor-type transport, which are controlled by
long-ranged momentum transport in the surrounding fluid.

The theory separates two issues that are often intertwined: the transport
mechanism that drives an interfacial instability and the modal structure
of the unstable branch. If the transverse orientational mode is damped,
it can be eliminated and the local theory reduces to a scalar height
equation with a renormalized active \(q^2\) coefficient. If this mode
remains soft, as in a symmetry-preserving two-dimensional
single-elastic-constant theory, the scalar reduction is not
asymptotically closed and the appropriate long-wavelength description is
a coupled height--director problem. In this case the instability may
remain local and diffuse-interface in origin even though the unstable
eigenmode is intrinsically mixed.

This framework identifies a diffuse-interface route to active interfacial
instability that can operate while the homogeneous active nematic remains
linearly stable because of hydrodynamic screening. It also provides a
basis for distinguishing local diffuse-interface instabilities,
bulk-flow-driven hydrodynamic instabilities, and mixed regimes in active
nematic--isotropic interfaces.
\end{abstract}

\maketitle

\section{Introduction}

Interfaces separating isotropic and nematic phases in active soft
materials display nonequilibrium instabilities whose physical origin can
differ qualitatively from that of passive interfaces. In a purely
diffusive passive binary mixture, the long-wavelength relaxation of a
conserved interfacial height field is controlled by capillarity and
diffusive transport. In an active nematic mixture, orientational order
generates active stresses~\cite{Ramaswamy2002,Ramaswamy2010} that can
feed back on the interface and destabilize an otherwise stable coexistence
profile~\cite{blow2014,gulati2024}.

A useful way of organizing such instabilities is to distinguish the
transport mechanism that communicates stresses to the interface from the
modal structure of the unstable branch. In a classical
Saffman--Taylor or Hele--Shaw setting, the interface is treated as a sharp 
moving boundary coupled to long-ranged incompressible flow. Eliminating the outer flow produces non-analytic contributions to the height dynamics, such as terms
proportional to \(|q|\) and \(|q|q^2\)~\cite{saffman1958,homsy1987}.
In active nematics, related hydrodynamic channels may arise either from
an externally imposed flow or from spontaneous active currents in the
ordered phase. The latter are associated with the bulk bend or splay
instability of the homogeneous active nematic~\cite{Ramaswamy2002,Marchetti2013},
and, at larger activity, with nonlinear active-turbulent
states~\cite{Wensink14308,Thampi2013PRL,Alert2022b}. In a screened diffuse-interface regime, by
contrast, the interface is not imposed as a sharp boundary. It is the finite-width region
over which the order parameter varies, as in Cahn--Hilliard/Model-B descriptions~\cite{Cahn1958,Hohenberg1977}. Projecting the linearised bulk dynamics onto the interfacial modes gives an interfacial stability operator with
a regular small-$q$ expansion, for finite hydrodynamic screening length~\cite{Kawasaki1982,TurskiLanger1980,JasnowZia1987}.

Passive nematic--isotropic interfaces are already non-scalar~\cite{deGennes1971} dynamical
objects. In addition to the interfacial height, orientational degrees of
freedom in the nematic phase can couple to surface modes and modify their
dispersion relation~\cite{popanita2003,popanita2005}; see also
Refs.~\cite{Schmid2007,Coelho2026} for a broader discussion of
fluctuating nematic--isotropic interfaces. Activity does not create this
multicomponent character from scratch; rather, it modifies the couplings
between height, scalar profile distortion or amplitude, and transverse orientational modes and can
turn them into growing or propagating interfacial disturbances. Recent
work on active liquid-crystal/passive-fluid interfaces has shown that
traveling capillary waves~\cite{adkins2022} can be described by a minimal theory in which the interfacial height is coupled non-reciprocally to the nematic
director at the interface~\cite{gulati2024}.

Active interfaces have also been studied in related, but distinct,
continuum settings. Numerical simulations of biphasic active nematics
have shown that activity can generate effective anchoring at
nematic--isotropic interfaces and produce undulatory interfacial
instabilities~\cite{blow2014,Giomi2014,Coelho2020,Coelho2021ptrsa,Coelho2023}.  At a more minimal
level, scalar active phase-field theories, including Active Model B and
Active Model B+, show that activity can modify coexistence, interfacial
fluctuations, and the effective capillary tension even in the absence of
orientational order~\cite{TjhungPRX2018,Fausti2021,catesnardini2025}.
When coupled to hydrodynamics, scalar active models can also exhibit
activity-induced interfacial stresses and arrested
coarsening~\cite{Tiribocchi2015}. These works establish that activity can
qualitatively alter interfacial dynamics, but they do not address the
long-wavelength modal structure of a diffuse nematic--isotropic
interface, where a conserved translational mode is coupled to amplitude
and transverse orientational modes of a tensorial order parameter.

A closely related analytical treatment is the fingering instability of
active nematic droplets studied by Alert~\cite{alert2022}. In that
sharp-interface model, nematic order is induced by anchoring at the
droplet boundary and is assumed to relax instantaneously, so that the
interface position is the only dynamical field. Shape perturbations distort the
slaved nematic texture and generate active flows that amplify the
deformation, even in the absence of a bulk active-nematic instability.

The present work differs from previous ones in two respects. First, the 
interface is a finite-width Cahn--Hilliard--Landau--de Gennes coexistence profile. In this sense the calculation is related to earlier interfacial-mode and collective-coordinate descriptions of diffuse interfaces~\cite{Kawasaki1982,TurskiLanger1980,JasnowZia1987}. However, the present  reduction is not closed on a single height field: the
translational mode is projected together with the amplitude and the
transverse-orientational modes. 
It also differs from recent sharp-interface reductions used
for active nematic interfaces~\cite{alert2022,gulati2024}. Here the projected long-wavelength operator remains tied to the finite-width interfacial structure and to the tensorial nematic order parameter, and contains both a direct active height contribution and possible height--director couplings.

\begin{figure*}
\centering
\includegraphics[width=0.8\linewidth]{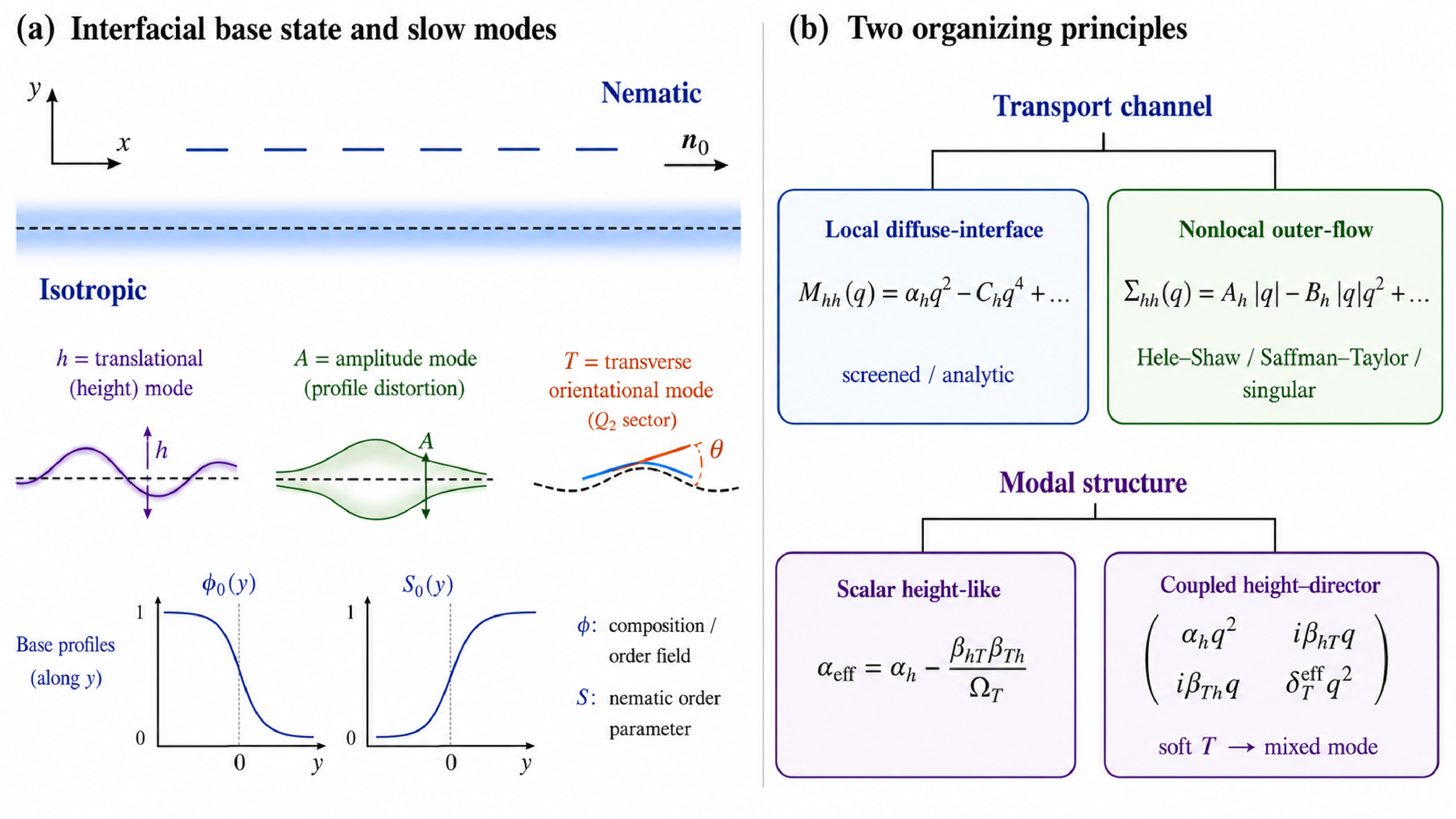}
\caption{
Diffuse-interface framework for active nematic--isotropic interfaces.
(a) Stationary planar base state with coexistence profiles
\(\phi_0(y)\) and \(S_0(y)\), and the three projected interfacial
degrees of freedom retained in the reduced description: the conserved
height mode \(h\), a scalar profile-distortion or amplitude mode \(A\), and
the transverse orientational mode \(T\). The sketch
of \(A\) represents a typical localized scalar profile distortion, here
drawn as a width-like or breathing deformation, rather than a uniquely
defined eigenfunction. The transverse mode
corresponds to a small director angle \(\theta\), measured from the base
direction \(\mathbf n_0=\hat{\mathbf x}\), with
\(Q_2=S\sin 2\theta\simeq 2S_0\theta\).
(b) Two organizing principles for interfacial instabilities: the
transport channel, distinguishing local analytic diffuse-interface
contributions
\(\mathcal{M}_{hh}(q)=\alpha_h q^2-C_h q^4+\cdots\)
from nonlocal singular outer-flow terms
\(\Sigma_{hh}(q)=A_h|q|-B_h|q|q^2+\cdots\), and the modal structure,
distinguishing a scalar height-like regime from a coupled
height--director regime with off-diagonal couplings
\(i\beta_{hT}q\) and \(i\beta_{Th}q\).
}
\label{figure1}
\end{figure*}

In what follows, we consider a binary active nematic mixture described by
a conserved composition field \(\phi\) and a two-dimensional nematic
tensor order parameter \(Q\), coupled to screened Stokes--Brinkman
hydrodynamics through elastic, capillary, and active
stresses~\cite{Coelho2026}. Linearizing about a flat nematic--isotropic
interface, we project the dynamics onto three slow degrees of freedom:
the translational mode associated with the conserved height \(h\), a damped amplitude mode \(A\), and a transverse orientational mode
\(T\). Eliminating the amplitude sector gives a reduced
long-wavelength operator for the coupled \((h,T)\) dynamics (see Fig.~\ref{figure1}).

The central result is that, in the screened diffuse-interface regime,
activity generates a direct local analytic contribution proportional to
\(q^2\) in the height sector of the projected interfacial stability operator, ${\cal M}$,
\begin{equation}
{\cal M}_{hh}(q)
=
\alpha_h q^2 - C_h q^4+\cdots .
\label{eq:intro_height_sector}
\end{equation}
The \(q^4\) term is the passive local diffusive capillary relaxation of
the conserved height mode~\cite{Cahn1958,Hohenberg1977,Kawasaki1982,TurskiLanger1980}. The \(q^2\) term is active and local: it arises by eliminating the screened flow generated by active stresses and projecting the resulting dynamics onto the translational mode of the diffuse interface.
In the sign convention
used here, the bare coefficient has the form
\begin{equation}
\alpha_h \sim \zeta (S_N+\xi) I_h,
\qquad
I_h>0 ,
\label{eq:intro_alpha_sign}
\end{equation}
so that it is destabilizing for extensile ($\zeta>0$), rod-like flow-aligning $(\xi>0)$ active nematics (order parameter $S_N>0$). The numerical value of \(I_h\) is not universal; it depends on the
interfacial profiles, the screened hydrodynamic Green function, friction,
viscosity, and boundary geometry. This local \(q^2\) height-sector contribution should be distinguished from
the singular height-sector terms generated by Saffman--Taylor/Hele--Shaw
outer-flow dynamics, and from the corresponding height-sector terms used in
recent active liquid-crystal/passive-fluid interface models~\cite{saffman1958,homsy1987,adkins2022,gulati2024}.

The direct coefficient \(\alpha_h\) is not, however, the only ingredient
in the long-wavelength theory. Translation invariance and reflection symmetry along the interface imply that the leading height--transverse couplings in {$\cal M$} are odd in \(q\),
\begin{equation}
{\cal M}_{hT}(q)=i\beta_{hT}q+\cdots,
\qquad
{\cal M}_{Th}(q)=i\beta_{Th}q+\cdots .
\label{eq:intro_ht_couplings}
\end{equation}
The gapped amplitude mode renormalizes the passive \(q^4\) height
stiffness but does not renormalize the direct active \(q^2\) coefficient.
The transverse orientational mode is different. If it is effectively damped, with
finite relaxation rate \(\Omega_T>0\), it can be eliminated and the
height-like growth rate is controlled by
\begin{equation}
\alpha_{\rm eff}
=
\alpha_h
-
\frac{\beta_{hT}\beta_{Th}}{\Omega_T}.
\label{eq:intro_alpha_eff}
\end{equation}
Thus \(\alpha_h\) is the bare local active height-channel contribution,
whereas \(\alpha_{\rm eff}\) is the scalar height coefficient after
transverse orientational feedback has been included.

If the transverse orientational mode is soft, the scalar height reduction
is not asymptotically closed. This occurs naturally in a strictly
two-dimensional single-elastic-constant theory without explicit
anchoring, elastic anisotropy, or finite-size symmetry breaking. The
appropriate local long-wavelength theory is then given by the coupled $(h,T)$ projected operator
\begin{equation}
{\cal M}^{(hT)}(q)
=
\begin{pmatrix}
\alpha_h q^2-C_h q^4
&
i\beta_{hT}q
\\
i\beta_{Th}q
&
-\Omega_T+\delta_T^{\rm eff}q^2
\end{pmatrix}
+\cdots ,
\label{eq:intro_hT_matrix}
\end{equation}
with \(\Omega_T=0\) in the rotationally invariant limit. In this regime
the unstable branch, when present, is generically a mixed
height--director mode. The instability may still be local and
diffuse-interface in origin even though the eigenmode is not a pure
height fluctuation. Coupled height--director descriptions also appear in
passive nematic--isotropic surface-wave theories and in recent
active-interface models~\cite{popanita2005,adkins2022,gulati2024}.

This separation between transport mechanism and modal structure is the
main organizing principle of the paper.  The analytic structure of the
projected operator identifies the transport channel: local diffuse-interface
transport gives regular powers of \(q\), while weakly screened outer-flow
transport gives non-analytic terms such as \(|q|\) and \(|q|q^2\).  The
modal structure is controlled separately by the transverse orientational
sector, which determines whether the unstable branch is height-like or
intrinsically mixed.

Because hydrodynamic screening shifts the
bulk bend or splay instability to a finite activity
threshold~\cite{Thampi2014,Doostmohammadi2018}, the homogeneous active
nematic may remain linearly stable even when the interface is
destabilized by the projected local active channel, providing a route
to active interfacial instability below the bulk active-nematic
threshold.

The aim of the present work is not to evaluate all projection
coefficients for a specific parameter set. Instead, we derive the
long-wavelength structure of the interfacial stability operator and
identify the terms that distinguish local diffuse-interface transport
from singular hydrodynamic transport, and scalar from coupled modal
regimes. The coefficients are therefore left in projected form, apart
from unbounded screened-kernel expressions that make their physical
origin explicit. Their numerical evaluation for particular material
parameters, boundary conditions, and geometries is a separate step,
required for quantitative stability diagrams and direct comparison with
experiments.

The paper is organized as follows. Section~II introduces the
Cahn--Hilliard--Landau--de Gennes model, the Stokes--Brinkman
hydrodynamic coupling, and the bulk transverse mode whose softening
controls the validity of a scalar interfacial reduction. Section~III
derives the projected diffuse-interface theory and reduces the three-mode
\((h,A,T)\) operator to the coupled \((h,T)\) sector. Section~IV shows how
the local analytic operator is supplemented by singular height-sector
terms in weakly screened Hele--Shaw/Saffman--Taylor-type geometries.
Section~V summarizes the resulting classification of transport channels
and modal regimes. Technical details of the mode construction, projection
procedure, and coefficient definitions are collected in the appendices.

\section{Model and Bulk Stability}

The purpose of this section is twofold. First, it fixes the notation and
sign conventions used in the interfacial projection. 
Second, it identifies the transverse bulk mode and the screened bulk
activity threshold that provide the reference point for distinguishing
interfacial from bulk-driven instabilities.

\subsection{Free energy functional}
We consider a binary mixture of an isotropic fluid and an active nematic, 
described by a conserved scalar composition field $\phi(\mathbf r,t)$ and a 
symmetric traceless nematic order parameter tensor $Q_{ij}(\mathbf r,t)$. 
The system is governed by a Cahn--Hilliard--Landau--de~Gennes free energy~\cite{Cahn1958,Hohenberg1977,deGennes1993} and 
Beris--Edwards hydrodynamics~\cite{beris1994thermodynamics} supplemented by an active stress of the form 
$-\zeta Q_{ij}$~\cite{Ramaswamy2002}. The free energy is given by
\begin{equation}
F[\phi,{\rm Q}] = \int d^d r \,
\Big[ f_{\rm loc}(\phi,{\rm Q}) + f_{\rm grad}(\phi,{\rm Q}) \Big],
\end{equation}
with local and gradient contributions
\begin{align}
f_{\rm loc}(\phi,{\rm Q}) &=
\frac{A}{2}\phi^2 + \frac{B}{4}\phi^4
+ \frac{a}{2}{\rm Tr(Q^2)} \nonumber \\
& - \frac{b}{3}{\rm Tr(Q^3)}
+ \frac{c}{4}\big[ \rm Tr (Q^2)\big]^2
- \lambda \phi \rm Tr(Q^2), \label{floc}\\
f_{\rm grad}(\phi,\rm Q) &=
\frac{\kappa}{2}|\nabla\phi|^2
+ \frac{L}{2}(\partial_k Q_{ij})(\partial_k Q_{ij}). \label{fgrad}
\end{align}
The coupling constant $\lambda>0$ favours nematic order in regions of large 
composition $\phi$. The parameters $\kappa$ and $L$ control, respectively, the 
interfacial stiffness of the composition field and the one--constant elastic 
penalty of the nematic. The remaining material dependent coefficients set the phase behaviour of the mixture, details of which may be found in~\cite{Coelho2026}.

The chemical potential associated with the conserved field $\phi$ is
\begin{equation}
\mu \equiv \frac{\delta F}{\delta\phi}
= A\phi + B\phi^3 - \lambda {\rm Tr(Q^2)} - \kappa \nabla^2 \phi .
\end{equation}

The ``raw'' molecular field conjugate to $Q_{ij}$ is
\begin{align}
G_{ij} &\equiv \frac{\delta F}{\delta Q_{ij}}
= a Q_{ij}
- b ({\rm Q}^2)_{ij}
+ c {\rm Tr(Q^2)} Q_{ij} \nonumber \\ &
- 2\lambda \phi Q_{ij}
- L \nabla^2 Q_{ij},
\end{align}
where $({\rm Q}^2)_{ij}=Q_{ik}Q_{kj}$. The traceless molecular field entering the 
Beris--Edwards equation is
\begin{equation}
H_{ij} = -G_{ij} + \frac{\delta_{ij}}{d}{\rm Tr G} ,
\end{equation}
with $d$ the spatial dimension.

\subsection{Dynamics, Hydrodynamics and Stresses}

The composition field obeys an advected Cahn--Hilliard equation~\cite{Cahn1958,Cahn1961},
\begin{equation}
\partial_t \phi + \mathbf v \cdot \nabla \phi
= \nabla \cdot \left( M_\phi \nabla \mu \right),
\label{CH}
\end{equation}
where $M_\phi$ is the mobility and $\mathbf v$ the velocity field.

The nematic tensor evolves according to the Beris--Edwards equation~\cite{beris1994thermodynamics},
\begin{equation}
(\partial_t + v_k \partial_k) Q_{ij} - S_{ij}
= \Gamma_Q H_{ij},
\label{BE}
\end{equation}
with rotational mobility $\Gamma_Q$ and co-rotational term
\begin{align}
S_{ij} &= (\xi A_{ik} + W_{ik})\left(Q_{kj}+\frac{\delta_{kj}}{d}\right) \nonumber\\
&
       + \left(Q_{ik}+\frac{\delta_{ik}}{d}\right)(\xi A_{kj}-W_{kj})
\nonumber\\
&\quad
- 2\xi \left(Q_{ij}+\frac{\delta_{ij}}{d}\right) Q_{kl}A_{kl}.
\label{Sij}
\end{align}
Here $A_{ij}$ and $W_{ij}$ are the strain-rate and vorticity tensors,
\begin{equation}
A_{ij}=\frac{1}{2}(\partial_j v_i+\partial_i v_j),
\qquad
W_{ij}=\frac{1}{2}(\partial_j v_i-\partial_i v_j),
\end{equation}
and $\xi$ is the flow-alignment parameter. We adopt the convention that
$\xi>0$ corresponds to rod-like particles. 

We assume incompressibility,
\begin{equation}
\nabla \cdot \mathbf v = 0,
\end{equation}
and use a Brinkman--Stokes force balance,
\begin{equation}
-\nabla p + \eta \nabla^2 \mathbf v - \Gamma \mathbf v
+ \nabla \cdot \left( \boldsymbol{\sigma}^{\rm el}
+ \boldsymbol{\sigma}^{\rm cap}
+ \boldsymbol{\sigma}^{\rm act} \right) = 0 ,
\label{Brinkman}
\end{equation}
where $\eta$ is the shear viscosity and $\Gamma$ a friction coefficient, with $\ell_B=\kappa_0^{-1}=\sqrt{\eta/\Gamma}$, the Brinkman screening length. The limits $\Gamma=0$ and $\eta\to 0$ correspond respectively to the Stokes
(wet) and Darcy (dry) regimes.

The elastic (Beris--Edwards) stress may be written as
\begin{align}
\sigma^{\rm el}_{ij} &=
2\xi \left(Q_{ij}+\frac{\delta_{ij}}{d}\right)Q_{kl}H_{kl}
- \xi H_{ik}\left(Q_{kj}+\frac{\delta_{kj}}{d}\right) \nonumber\\
&
- \xi \left(Q_{ik}+\frac{\delta_{ik}}{d}\right)H_{kj} - \frac{\partial f}{\partial(\partial_j Q_{k\ell})}\,\partial_i Q_{k\ell}
\nonumber\\
&\quad
+ Q_{ik}H_{kj} - H_{ik}Q_{kj}.
\end{align}

The anisotropic capillary (Korteweg) stress associated with composition
gradients is
\begin{equation}
\sigma^{\rm cap}_{ij} = -\kappa\, \partial_i \phi\, \partial_j \phi,
\end{equation}
while the isotropic thermodynamic contributions are absorbed into the pressure $p$.

Activity is introduced through the standard active stress~\cite{Ramaswamy2002}
\begin{equation}
\sigma^{\rm act}_{ij} = -\zeta Q_{ij}.
\end{equation}
We adopt the convention that \(\zeta>0\) corresponds to extensile
activity.

Although we started from the standard tensorial notation~\cite{Coelho2026}, the analysis carried out throughout this paper is strictly two-dimensional and will be rewritten in
terms of the two independent components $(Q_1,Q_2)$ below.

\subsection{Two-dimensional representation}

In two dimensions the symmetric traceless tensor $Q_{ij}$ can be written in
terms of two independent components,
\begin{equation}
Q_{ij}=\frac{1}{2}
\begin{pmatrix}
Q_1 & Q_2\\
Q_2 & -Q_1
\end{pmatrix},
\quad
Q_1 = Q_{xx}-Q_{yy}, \quad Q_2 = 2Q_{xy}.
\end{equation}
The invariants reduce to
\begin{align}
&{\rm Tr (Q}^2)=\frac{1}{2}(Q_1^2+Q_2^2)=\frac{1}{2} S^2, \nonumber \\ &
(\partial_k Q_{ij})(\partial_k Q_{ij})
=\frac{1}{2}\left(|\nabla Q_1|^2+|\nabla Q_2|^2\right).
\end{align}
where $S$ is the usual nematic scalar order parameter.

For a strictly two-dimensional symmetric traceless tensor,
\(\operatorname{Tr}(Q^3)=0\). The cubic invariant in the tensorial
three-dimensional free energy therefore drops out of the two-dimensional
representation used in the remainder of the paper. The free-energy density becomes
\begin{align}
&f(\phi,Q_1,Q_2)=
\frac{A}{2}\phi^2+\frac{B}{4}\phi^4
+\frac{\kappa}{2}|\nabla\phi|^2
+(\frac{a}{4}-\frac{\lambda}{2}\phi)
(Q_1^2 \nonumber \\ & +Q_2^2) 
+\frac{c}{16}(Q_1^2+Q_2^2)^2
+\frac{L}{4}\left(|\nabla Q_1|^2+|\nabla Q_2|^2\right).
\end{align}

Defining
\begin{equation}
m(\phi,Q_1,Q_2)=\frac{a}{2}
+\frac{c}{4}(Q_1^2+Q_2^2)-\lambda\phi ,
\end{equation}
the molecular fields read
\begin{equation}
H_1 = -m Q_1 + \frac{L}{2}\nabla^2 Q_1,
\qquad
H_2 = -m Q_2 + \frac{L}{2}\nabla^2 Q_2.
\end{equation}

Thus, in the absence of explicit anchoring or elastic anisotropy, the passive
two-dimensional theory is invariant under rotations in the \((Q_1,Q_2)\)
plane. A uniform rotation of a nematic state changes \(Q_1\) into \(Q_2\) at
no local free-energy cost. Thus the \(Q_2\) sector is the transverse rotational sector of the
two-dimensional theory.

Finally, the active stresses are
\begin{equation}
\sigma^{\rm act}_{xx}=-\frac{\zeta}{2}Q_1,\qquad
\sigma^{\rm act}_{yy}=+\frac{\zeta}{2}Q_1,\qquad
\sigma^{\rm act}_{xy}=-\frac{\zeta}{2}Q_2 .
\end{equation}

Equations \eqref{CH}, \eqref{BE}, and \eqref{Brinkman}, together with the above
definitions, constitute the starting point for the bulk and interfacial linear
stability analyses carried out in the following sections.

\subsection{Bulk linear modes in the aligned geometry}
\label{Bulk_linear_modes}

We now examine the linear bulk dynamics about a homogeneous nematic base
state in the aligned geometry used later for the interfacial problem. 
The aim is not a complete classification of bulk instabilities, but to
identify the transverse orientational sector that couples to the
interfacial problem and to fix the screened bulk threshold used below.

We consider a uniform nematic with concentration \(\phi_0\) and director
aligned along \(\hat{\bf x}\). In the two-dimensional representation,
\begin{equation}
Q_1=S_0,\qquad Q_2=0 ,
\end{equation}
where \(S_0\) is fixed by the homogeneous mean-field equation. We consider
longitudinal perturbations,
\begin{equation}
{\bf q}=q\hat{\bf x},\qquad {\bf n}_0=\hat{\bf x},
\end{equation}
of the form
\begin{equation}
(\delta\phi,\delta Q_1,\delta Q_2,\delta{\bf v})
\propto e^{\sigma t+iqx}.
\end{equation}

In the absence of flow and activity, the linearised Cahn--Hilliard and
Beris--Edwards equations are
\begin{equation}
\partial_t
\begin{pmatrix}
\delta\phi\\
\delta Q_1\\
\delta Q_2
\end{pmatrix}
=
{\cal L}^{(0)}(q)
\begin{pmatrix}
\delta\phi\\
\delta Q_1\\
\delta Q_2
\end{pmatrix},
\end{equation}
with
\begin{equation}
\scriptsize
\setlength{\arraycolsep}{2pt}
\begin{split}
&{\cal L}^{(0)}(q)=
\\
&
\begin{pmatrix}
-M_\phi q^2(f_{\phi\phi}+\kappa q^2)
&
-M_\phi q^2 f_{\phi Q_1}
&
0
\\
-\Gamma_Q f_{Q_1\phi}
&
-\Gamma_Q(f_{Q_1Q_1}+Lq^2/2)
&
0
\\
0
&
0
&
-\Gamma_Q(f_{Q_2Q_2}+Lq^2/2)
\end{pmatrix}.
\end{split}
\end{equation}
The local curvatures are
\begin{equation}
f_{\phi\phi}=A+3B\phi_0^2,\qquad
f_{\phi Q_1}=f_{Q_1\phi}=-\lambda S_0,
\end{equation}
\begin{equation}
f_{Q_1Q_1}=\frac{a}{2}+\frac{3c}{4}S_0^2-\lambda\phi_0,
\qquad
f_{Q_2Q_2}=\frac{a}{2}+\frac{c}{4}S_0^2-\lambda\phi_0 .
\end{equation}
In the ordered nematic phase, the mean-field condition for \(S_0\neq0\)
implies \(f_{Q_2Q_2}=0\). Thus the \(Q_2\) branch is the transverse
Goldstone sector of the passive homogeneous nematic, while
\((\delta\phi,\delta Q_1)\) form the coupled composition--amplitude
sector.

Eliminating the velocity field from the linearised Brinkman--Stokes
equation gives
\begin{equation}
(\eta q^2+\Gamma)\delta v_i
=
-\zeta P_{ij}(\hat{\bf q})\,iq_k\delta Q_{jk},
\qquad
P_{ij}=\delta_{ij}-\hat q_i\hat q_j .
\end{equation}
For \(q\parallel\hat{\bf x}\), incompressibility removes the longitudinal
velocity component and only the shear flow \(\delta v_y\) remains. This
flow is proportional to the transverse fluctuation \(\delta Q_2\), so the
active hydrodynamic feedback modifies only the \(Q_2\) equation. The full
bulk operator is
\begin{equation}
{\cal L}(q)
=
{\cal L}^{(0)}(q)+{\cal L}^{\rm act}(q),
\end{equation}
with
\begin{equation}
\label{eq:Lact-bulk}
{\cal L}^{\rm act}(q)
=
\begin{pmatrix}
0&0&0\\
0&0&0\\
0&0&
\dfrac{\zeta q^2}{2(\eta q^2+\Gamma)}(S_0+\xi)
\end{pmatrix}.
\end{equation}
This is the standard active-nematic feedback in the aligned transverse sector: a
transverse director distortion drives an active shear flow, and the
flow-alignment term feeds this flow back into the transverse dynamics.
For \(\Gamma>0\), the response is regular at small \(q\), and the active
correction is analytic in the long-wavelength limit.

The transverse branch has growth rate
\begin{equation}
\sigma_T(q)
=
-\Gamma_Q\left(f_{Q_2Q_2}+\frac{L}{2}q^2\right)
+
\frac{\zeta q^2}{2(\eta q^2+\Gamma)}(S_0+\xi).
\end{equation}
In the nematic phase, \(f_{Q_2Q_2}=0\). Expanding at small \(q\) for
\(\Gamma>0\) gives
\begin{equation}
\sigma_T(q)
=
q^2
\left[
-\frac{\Gamma_Q L}{2}
+
\frac{\zeta(S_0+\xi)}{2\Gamma}
\right]
+O(q^4).
\end{equation}
Thus substrate friction shifts the bulk bend instability~\cite{Ramaswamy2002,Marchetti2013} to the finite
threshold~\cite{Thampi2014,Doostmohammadi2018}
\begin{equation}
\zeta_c^{\rm bulk}
=
\frac{\Gamma\Gamma_Q L}{S_0+\xi},
\qquad S_0+\xi>0.
\end{equation}

More generally, if \(\theta\) is the angle between $\mathbf q$ and ${\bf n}_0$,
\begin{align}
&\sigma_\theta(q,\theta)
= \nonumber \\ &
-\Gamma_Q\left(f_{Q_2Q_2}+\frac{L}{2}q^2\right)
+
\frac{\zeta q^2}{2(\eta q^2+\Gamma)}
\left(\xi+S_0\cos2\theta\right).
\end{align}
For extensile activity, \(\zeta>0\), the maximum occurs at
\(\theta=0\). This is the aligned sector used in the interfacial
calculation.

The bulk analysis therefore supplies the reference threshold for the
interfacial problem. In the regime
\begin{equation}
\zeta<\zeta_c^{\rm bulk}(\Gamma),
\end{equation}
the homogeneous active nematic is linearly stable, so any instability of
the projected diffuse-interface operator is not bulk-driven in the sense
of a positive homogeneous bulk growth rate.

\section{Interfacial stability: diffusive interface regime}

\subsection{Geometry and interfacial base state}

We consider a two-dimensional system with coordinates $\mathbf r=(x,y)$ and a planar interface separating an isotropic phase ($y\to-\infty$) from a nematic phase ($y\to+\infty$). In the restricted geometry introduced above, the base state is translationally invariant along the interface and varies only across it, so that
\begin{equation}
\phi(\mathbf r,t)=\phi_0(y), \qquad
{Q_1}(\mathbf r,t)=S_0(y), \qquad
{Q_2}(\mathbf r,t)=0 .
\end{equation}
The nematic director is therefore uniform and aligned with the $x$-axis in the ordered phase. This is the interfacial analogue of the aligned bulk geometry discussed
in Sec.~II.

The planar interfacial equilibrium profiles $\phi_0(y)$ and $S_0(y)$ satisfy the one-dimensional Euler--Lagrange equations
\begin{equation}
\mu_0 = A\phi_0 + B\phi_0^3 - \frac{\lambda}{2} S_0^2 - \kappa \partial_y^2 \phi_0,
\label{eq:ELphi}
\end{equation}
\begin{equation}
0 = \left(a - 2\lambda \phi_0 + \frac{c}{2} S_0^2\right) S_0 - \frac{L}{2}\partial_y^2 S_0 ,
\label{eq:ELS}
\end{equation}
with boundary conditions
\begin{align}
&(\phi_0,S_0)\to(\phi_I,0)\quad (y\to-\infty),\nonumber \\ &
(\phi_0,S_0)\to(\phi_N,S_N)\quad (y\to+\infty),
\end{align}
where $(\phi_I,0)$ and $(\phi_N,S_N)$ are the coexisting bulk isotropic and nematic states.

\subsection{Linearised interfacial operator}

We consider small perturbations about the flat interfacial base state,
\begin{align}
&\phi=\phi_0(y)+\delta\phi(x,y,t), \nonumber \\ &
Q_1=S_0(y)+\delta Q_1(x,y,t), \nonumber \\ &
 Q_2 = \delta Q_2(x,y,t),\qquad 
\end{align}
and decompose them into Fourier modes along the interface,
\begin{equation}
(\delta\phi,\delta Q_1,\delta Q_2)(x,y,t)
=
(\phi_q,Q_{1q},Q_{2q})(y)\,e^{\sigma t+iqx}.
\end{equation}
Linearising the dynamical equations, \eqref{CH} and \eqref{BE}, about the inhomogeneous base profiles then yields an eigenvalue problem of the form
\begin{equation}
\sigma
\begin{pmatrix}
\phi_q\\
Q_{1q}\\
Q_{2q}
\end{pmatrix}
=
\mathcal{L}(q;y)
\begin{pmatrix}
\phi_q\\
Q_{1q}\\
Q_{2q}
\end{pmatrix}.
\end{equation}
where $\mathcal L(q;y)$ is a differential operator in $y$ obtained from the homogeneous bulk operator by evaluating its coefficients on the planar interfacial profiles $\phi_0(y)$ and $S_0(y)$.

In the absence of activity and hydrodynamic couplings, the linear operator reduces to the passive interfacial operator
\begin{equation}
\scriptsize
\setlength{\arraycolsep}{2pt}
\begin{split}
&\mathcal L^{(0)}(q;y)= \\
& \begin{pmatrix}
M_\phi(\partial_y^2-q^2)\,\bar f_{\phi\phi}(q,y) &
M_\phi(\partial_y^2-q^2)\,f_{\phi Q_1}(y) & 0 \\
-\Gamma_Q f_{Q_1\phi}(y) &
-\Gamma_Q \bar f_{Q_1Q_1}(q,y) & 0 \\
0 & 0 & -\Gamma_Q \bar f_{Q_2Q_2}(q,y)
\end{pmatrix},
\end{split}
\label{eq:passive_bulk_operator}
\end{equation}
with
\begin{align}
\bar f_{\phi\phi}(q,y) &= f_{\phi\phi}(y)-\kappa(\partial_y^2-q^2), \\
\bar f_{Q_1Q_1}(q,y) &= f_{Q_1Q_1}(y)-\frac{L}{2}(\partial_y^2-q^2), \\
\bar f_{Q_2Q_2}(q,y) &= f_{Q_2Q_2}(y)-\frac{L}{2}(\partial_y^2-q^2),
\end{align}
and
\begin{align}
f_{\phi\phi}(y) &= A+3B\phi_0^2(y), \\
f_{\phi Q_1}(y)&=f_{Q_1\phi}(y) = -\lambda S_0(y), \\
f_{Q_1Q_1}(y) &= \frac{a}{2}+\frac{3c}{4}S_0^2(y)-\lambda\phi_0(y), \\
f_{Q_2Q_2}(y) &= \frac{a}{2}+\frac{c}{4}S_0^2(y)-\lambda\phi_0(y).
\end{align}

As in the homogeneous bulk problem, the passive interfacial operator has a block structure: the composition and amplitude sectors $(\phi_q,Q_{1q})$ are coupled, whereas the transverse director fluctuation $Q_{2q}$ remains decoupled.

\subsection{Slow interfacial modes and symmetry}
\label{sec:slow_modes}

Translational invariance of the free energy implies the existence of a
Goldstone mode associated with rigid shifts of the planar equilibrium
interface. For \(q=0\), the linear operator therefore admits the zero
mode
\begin{equation}
\boldsymbol{\Psi}_h(y)
=
\begin{pmatrix}
\partial_y\phi_0(y)\\
\partial_yS_0(y)\\
0
\end{pmatrix},
\qquad
\sigma(q=0)=0 .
\label{eq:height_shape_main}
\end{equation}
The overall sign has been chosen so that the collective coordinate
multiplying \(\boldsymbol{\Psi}_h\) is the physical height defined below.

To identify this mode with a physical interface displacement, we define
the conserved excess composition
\begin{equation}
h(x,t)
=
\frac{1}{\Delta\phi}
\int_{-\infty}^{\infty}dy\,\delta\phi(x,y,t),
\qquad
\Delta\phi=\phi_N-\phi_I .
\label{eq:height_def_main}
\end{equation}
With this convention, a geometrical displacement \(\eta\) of the base
profiles in the positive \(y\)-direction is represented as
\begin{eqnarray}
\delta\phi&=&-\eta\,\Psi_h^\phi=-\eta\,\partial_y\phi_0,  \\ \nonumber
\qquad
\delta Q_1&=&-\eta\,\Psi_h^{Q_1}=-\eta\,\partial_y S_0, \\ \nonumber
\qquad
\delta Q_2&=&0 .
\label{eq:geometric_displacement}
\end{eqnarray}
Therefore the excess-composition height defined in
Eq.~\eqref{eq:height_def_main} is \(h=-\eta\) for such a geometrical
displacement. Conversely, a perturbation \(h\boldsymbol{\Psi}_h\) has
\(\int dy\,\delta\phi=h\Delta\phi\), and hence its coefficient is
precisely the conserved height variable \(h(x,t)\).
Thus the translational Goldstone mode is uniquely associated with the
height field \(h(x,t)\), while the sign of a geometrical displacement is
set separately by the orientation of the \(y\)-axis.

Beyond the translational mode, the interfacial operator generally
contains additional localized or slow modes describing internal
distortions of the diffuse interface. In the present geometry the
relevant localized non-Goldstone mode in the coupled \((\phi,Q_1)\)
sector is denoted by \(\boldsymbol{\Psi}_A(y)\); its  amplitude \(A(x,t)\)
describes distortion of the interfacial profile. This
mode describes fluctuations of the scalar nematic order and composition profiles which do not correspond to a rigid
translation.

The transverse \(Q_2\) sector is qualitatively different. In a strictly
two-dimensional passive theory with a single elastic constant and no
explicit orientational anchoring, a uniform rotation of the nematic
director costs no bulk free energy. The corresponding infinitesimal
rotation of the interfacial base state gives the vector-valued transverse
shape function
\begin{equation}
\boldsymbol{\Psi}_T(y)
=
\begin{pmatrix}
0\\
0\\
\psi_T(y)
\end{pmatrix}
\propto
\begin{pmatrix}
0\\
0\\
S_0(y)
\end{pmatrix}.
\label{eq:transverse_shape_main}
\end{equation}
Here \(\psi_T(y)\) denotes the scalar \(Q_2\)-component of the
transverse shape function. Because \(S_0(y)\) approaches its bulk
nematic value on the nematic side, this mode is not, in general, a
localized interfacial bound state in the infinite-domain limit. It is
nevertheless a slow orientational degree of freedom of the interfacial
problem and must be retained in the long-wavelength theory whenever
height distortions couple to transverse director rotations.

The long-wavelength dynamics is therefore controlled by three
slow collective coordinates: the height \(h\), associated with
\(\boldsymbol{\Psi}_h\); the amplitude \(A\), associated with
\(\boldsymbol{\Psi}_A\); and the transverse orientational field
\(T\), associated with \(\boldsymbol{\Psi}_T\). We therefore write
\begin{eqnarray}
\delta\mathbf U(x,y,t)
=
h(x,t)\boldsymbol{\Psi}_h(y)
+
A(x,t)\boldsymbol{\Psi}_A(y)
+ \nonumber \\ 
T(x,t)\boldsymbol{\Psi}_T(y)
+
\text{fast modes},
\label{eq:slow_mode_expansion_main}
\end{eqnarray}
where
\begin{equation}
\delta\mathbf U=(\delta\phi,\delta Q_1,\delta Q_2)^T .
\label{eq:deltaU_def_main}
\end{equation}

In a finite system, or in the presence of anchoring, elastic anisotropy,
or explicit orientational symmetry breaking, the transverse sector may
acquire a small gap \(\Omega_T\). In the symmetry-preserving limit
considered here, \(\Omega_T=0\) at \(q=0\).

\subsection{Projected three-mode operator}

We now project the linearised interfacial dynamics onto the 
subspace spanned by the translational, amplitude, and transverse
orientational shape functions introduced above. The construction follows
the standard diffuse-interface strategy of identifying the slow
collective coordinates from the coexistence profile and projecting the
bulk linearised dynamics onto these modes~\cite{Kawasaki1982,TurskiLanger1980,JasnowZia1987}.
Here the procedure is extended to a nematic--isotropic interface, where
the slow subspace contains not only the conserved translational mode but
also amplitude and transverse orientational degrees of freedom.

Starting from
\begin{eqnarray}
\partial_t \delta\mathbf U_q(y,t)
&=&
{\cal L}(q;y)\,\delta\mathbf U_q(y,t), \nonumber \\
\qquad
\delta\mathbf U_q
&=&
(\delta\phi_q,\delta Q_{1q},\delta Q_{2q})^T ,
\label{eq:linear_operator_interfacial}
\end{eqnarray}
with \({\cal L}(q;y)\) given in the passive case by Eq.~\eqref{eq:passive_bulk_operator}
and supplemented, in the active system, by the corresponding
hydrodynamic correction evaluated on the interfacial background, we
restrict the dynamics to the slow interfacial sector by writing
\begin{equation}
\delta\mathbf U_q(y,t)
=
\sum_{i=h,A,T} u_i(q,t)\,\boldsymbol{\Psi}_i(y),
\qquad
u_i=(h_q,A_q,T_q).
\label{eq:slow_projection_main}
\end{equation}
Here
\(\boldsymbol{\Psi}_h\), \(\boldsymbol{\Psi}_A\), and
\(\boldsymbol{\Psi}_T\) are vector-valued shape functions in the
\((\phi,Q_1,Q_2)\) space, as defined in Sec.~\ref{sec:slow_modes}
and Appendix~\ref{app:shape_functions}.

Projection of the full operator onto this basis gives
\begin{equation}
\partial_t u_i(q,t)
=
\sum_{j=h,A,T}
{\cal M}_{ij}(q)\,u_j(q,t),
\label{eq:projected_dynamics_components}
\end{equation}
where the projected interfacial stability matrix is, in the orthogonal
basis used here,
\begin{equation}
{\cal M}_{ij}(q)
=
N_i^{-1}
\left\langle
\boldsymbol{\Psi}_i
\middle|
{\cal L}(q)
\middle|
\boldsymbol{\Psi}_j
\right\rangle,
\qquad
N_i=
\left\langle
\boldsymbol{\Psi}_i
\middle|
\boldsymbol{\Psi}_i
\right\rangle .
\label{eq:projected_matrix_main}
\end{equation}
The projection inner product, defined in Appendix~\ref{app:projection}, accounts 
for the different kinetic structures of the conserved composition field and the
relaxational nematic fields: the \(\phi\) sector is weighted by the inverse Cahn--Hilliard operator, while the \(Q\) sectors are weighted by $1/\Gamma_Q$.

With this choice, the normalization of
the translational mode is consistent with the conserved definition of
height in Eq.~\eqref{eq:height_def_main}, and the orthogonality
relations used below follow from the projection metric, as summarized in
Appendix~\ref{app:projection}.

Equation~\eqref{eq:projected_matrix_main} defines the wavevector-dependent
projected operator \({\cal M}(q)\) governing the linear dynamics of the
interfacial degrees of freedom. Its small-\(q\) structure is constrained
by conservation and symmetry, as discussed next.

\subsection{Symmetry-constrained three-mode matrix}

We now constrain the small-\(q\) structure of the projected interfacial
operator \({\cal M}(q)\). The slow interfacial variables are the height
\(h\), associated with the translational shape function
\(\boldsymbol{\Psi}_h\), the scalar amplitude \(A\), associated with
\(\boldsymbol{\Psi}_A\), and the transverse orientational amplitude
\(T\), associated with \(\boldsymbol{\Psi}_T\).

At zero longitudinal wavenumber, a uniform displacement of the interface
is a neutral mode. Moreover, with the amplitude and transverse modes
chosen independently of the translational mode, a spatially uniform
perturbation of \(A\) or \(T\) cannot produce a change of the conserved
interfacial height. Thus the height row of the projected operator
vanishes at \(q=0\),
\begin{equation}
{\cal M}_{hj}(0)=0,
\qquad
j=h,A,T .
\label{eq:height_row_constraint}
\end{equation}
This is the only general constraint on the height row that we impose at
this stage. The stronger diffusive factor \(q^2\) applies to the passive
Cahn--Hilliard height relaxation, but not to the active advective
contribution generated by hydrodynamic elimination.

Reflection symmetry along the interface, \(x\mapsto -x\), further
constrains the parity of the projected matrix elements. The height
coordinate \(h\) and the scalar amplitude \(A\) are even under this
reflection, whereas the transverse orientational amplitude \(T\) is odd,
because it is associated with the \(Q_2\) sector. We therefore assign
\begin{equation}
p_h=p_A=+1,
\qquad
p_T=-1 .
\label{eq:mode_parities}
\end{equation}
The projected operator satisfies
\begin{equation}
{\cal M}_{ij}(q)
=
p_i p_j\,{\cal M}_{ij}(-q).
\label{eq:parity_constraint}
\end{equation}
Thus entries coupling variables of the same parity are even functions of
\(q\), while entries coupling variables of opposite parity are odd
functions of \(q\). In Fourier representation, the leading odd terms are
purely imaginary and proportional to \(iq\).

Combining Eq.~\eqref{eq:height_row_constraint} with the parity
constraint gives the leading height-row structure
\begin{equation}
{\cal M}_{hh}=O(q^2),
\qquad
{\cal M}_{hA}=O(q^2),
\qquad
{\cal M}_{hT}=O(iq).
\label{eq:height_row_power_counting}
\end{equation}
The first two entries are even in \(q\) and vanish at \(q=0\), whereas
the height--tilt entry is odd in \(q\) and therefore begins linearly in
\(iq\).

The nonconserved sectors are not constrained in the same way. The scalar
amplitude mode is generally gapped, so that
\begin{equation}
{\cal M}_{AA}=-\Omega_A+O(q^2),
\qquad
\Omega_A>0 .
\label{eq:amplitude_gap_main}
\end{equation}
The transverse orientational mode may also be gapped by finite-size
effects, anchoring, elastic anisotropy, or explicit orientational
symmetry-breaking terms, in which case
\begin{equation}
{\cal M}_{TT}=-\Omega_T+O(q^2).
\label{eq:transverse_gap_main}
\end{equation}
In the symmetry-preserving two-dimensional theory with a single elastic
constant, however, the transverse mode is the orientational Goldstone
mode and \(\Omega_T=0\).

In the equilibrium diffuse-interface problem, detailed balance and
composition conservation further restrict the height dynamics: the
diagonal height relaxation is the diffuse capillary mode and begins at
order \(q^4\). Activity can add a local analytic \(q^2\) contribution to
this diagonal entry. In addition, active advection allows the leading
height--tilt coupling in Eq.~\eqref{eq:height_row_power_counting} to be
present.

We therefore write the leading long-wavelength interfacial matrix as
\begin{align}
\begin{split}
&{\cal M}(q) = \\
 & \begin{pmatrix}
&\alpha_h q^2-M_{\rm eff}\gamma q^4
&
\beta_{hA}q^2
&
i\beta_{hT}q
\\[2mm]
\beta_{Ah}q^2
&
-\Omega_A+\delta_A q^2
&
i\beta_{AT}q
\\[2mm]
i\beta_{Th}q
&
i\beta_{TA}q
&
-\Omega_T+\delta_T q^2
\end{pmatrix}
+\cdots .
\end{split}
\label{eq:three_mode_matrix}
\end{align}
Here \(\alpha_h\) denotes the direct active local contribution to the
conserved height dynamics, while \(M_{\rm eff}\gamma\) is the passive
capillary coefficient. The coefficient \(\beta_{hT}\) denotes the
leading retained height--tilt coupling in the height row; in the active
diffuse-interface channel emphasized below, it receives an advective
contribution generated by the hydrodynamic projection. The remaining
coefficients \(\beta_{ij}\) denote the leading symmetry-allowed
couplings between height, amplitude, and transverse orientational
variables, and \(\delta_A,\delta_T\) are the leading \(q^2\) corrections
in the nonconserved sectors.

Equation~\eqref{eq:three_mode_matrix} is the minimal analytic
long-wavelength matrix consistent with reflection symmetry, height
conservation, the passive diffuse-interface structure, and the active
hydrodynamic projection.

\subsection{Reduction to the coupled {$(h,T)$} sector}
\label{sec:hT_reduction}

The amplitude coordinate \(A\), associated with the localized shape
function \(\boldsymbol{\Psi}_A\), is assumed to remain gapped,
\(\Omega_A>0\), and can therefore be eliminated
perturbatively at long wavelengths. We reorder the basis as
\begin{equation}
{\bf u}
=
\begin{pmatrix}
h\\
T\\
A
\end{pmatrix},
\label{eq:reordered_basis_hTA}
\end{equation}
so that the slow sector appears first. The linearized equations then
have the block form
\begin{align}
&\partial_t
\begin{pmatrix}
h\\
T
\end{pmatrix}
=
{\cal M}_{ss}
\begin{pmatrix}
h\\
T
\end{pmatrix}
+
{\cal M}_{sA}A,
\nonumber \\ &
\partial_t A
=
{\cal M}_{As}
\begin{pmatrix}
h \\
T
\end{pmatrix}
+
{\cal M}_{AA}A,
\label{eq:block_form_hTA}
\end{align}
where \(s\) labels the slow \((h,T)\) sector. Thus
\({\cal M}_{ss}\) is a \(2\times 2\) block, \({\cal M}_{sA}\) is a
\(2\times 1\) block, \({\cal M}_{As}\) is a \(1\times 2\) block, and
\({\cal M}_{AA}\) is the scalar block acting on the amplitude mode.

When the amplitude mode relaxes rapidly compared with the long-wavelength
dynamics of \(h\) and \(T\), it is quasi-statically slaved to the slow
variables. Neglecting \(\partial_t A\) to leading order gives
\begin{equation}
A
=
-{\cal M}_{AA}^{-1}{\cal M}_{As}
\begin{pmatrix}
h\\
T
\end{pmatrix}.
\label{eq:A_slaving}
\end{equation}
Substitution into the slow equations yields the reduced operator
\begin{align}
&\partial_t
\begin{pmatrix}
h\\
T
\end{pmatrix}
=
{\cal M}^{(hT)}(q)
\begin{pmatrix}
h\\
T
\end{pmatrix},
\nonumber \\ &
{\cal M}^{(hT)}(q)
=
{\cal M}_{ss}
-
{\cal M}_{sA}{\cal M}_{AA}^{-1}{\cal M}_{As}.
\label{eq:Schur_hT}
\end{align}
This is the Schur complement of the fast amplitude sector. The reduction
does not require the transverse mode to be gapped; it only assumes that
the amplitude mode remains fast compared with the slow \((h,T)\)
dynamics.

We now insert the long-wavelength expansion of the projected
three-mode matrix into Eq.~\eqref{eq:Schur_hT}. In the reordered
\((h,T,A)\) basis, the leading blocks are
\begin{align}
{\cal M}_{ss}
&=
\begin{pmatrix}
\alpha_h q^2-M_{\rm eff}\gamma q^4
&
i\beta_{hT}q
\\[3pt]
i\beta_{Th}q
&
-\Omega_T+\delta_T q^2
\end{pmatrix}
+\cdots ,
\label{eq:Mss_expansion}
\\
{\cal M}_{sA}
&=
\begin{pmatrix}
\beta_{hA}q^2
\\[3pt]
i\beta_{TA}q
\end{pmatrix}
+\cdots ,
\,
{\cal M}_{As}
=
\begin{pmatrix}
\beta_{Ah}q^2
&
i\beta_{AT}q
\end{pmatrix}
+\cdots ,
\label{eq:MsA_MAs_expansion}
\\
{\cal M}_{AA}
&=
-\Omega_A+\delta_A q^2+\cdots .
\label{eq:MAA_expansion}
\end{align}
Since
\begin{equation}
{\cal M}_{AA}^{-1}
=
-\frac{1}{\Omega_A}
+O(q^2),
\label{eq:MAA_inverse_expansion}
\end{equation}
the Schur correction is, to the required order,
\begin{equation}
-{\cal M}_{sA}{\cal M}_{AA}^{-1}{\cal M}_{As}
=
\frac{1}{\Omega_A}
{\cal M}_{sA}{\cal M}_{As}
+\cdots .
\label{eq:Schur_correction_expansion}
\end{equation}

The height-height correction scales as
\begin{equation}
{\cal M}_{hA}{\cal M}_{AA}^{-1}{\cal M}_{Ah}
\sim q^2 q^0 q^2
\sim q^4,
\label{eq:hh_Schur_scaling}
\end{equation}
and therefore renormalizes the capillary coefficient but not the direct
active \(q^2\) height coefficient \(\alpha_h\). The transverse diagonal
correction scales as
\begin{equation}
{\cal M}_{TA}{\cal M}_{AA}^{-1}{\cal M}_{AT}
\sim q q^0 q
\sim q^2,
\label{eq:TT_Schur_scaling}
\end{equation}
and renormalizes the \(q^2\) stiffness of the transverse sector.

The off-diagonal corrections are subleading:
\begin{align}
&{\cal M}_{hA}{\cal M}_{AA}^{-1}{\cal M}_{AT}
\sim q^2 q^0 q
\sim q^3,
\nonumber \\ &
{\cal M}_{TA}{\cal M}_{AA}^{-1}{\cal M}_{Ah}
\sim q q^0 q^2
\sim q^3.
\label{eq:offdiag_Schur_scaling}
\end{align}
Thus elimination of the amplitude mode does not modify the leading
\(iq\) height--tilt couplings. It only generates subleading \(iq^3\)
corrections to the off-diagonal entries.

The reduced long-wavelength operator is therefore
\begin{equation}
{\cal M}^{(hT)}(q)
=
\begin{pmatrix}
\alpha_h q^2-C_h q^4
&
i\beta_{hT}q
\\[3pt]
i\beta_{Th}q
&
-\Omega_T+\delta_T^{\rm eff}q^2
\end{pmatrix}
+\cdots ,
\label{eq:reduced_hT_matrix}
\end{equation}
where
\begin{equation}
C_h
=
M_{\rm eff}\gamma
-
\frac{\beta_{hA}\beta_{Ah}}{\Omega_A},
\label{eq:Ch_definition}
\end{equation}
and
\begin{equation}
\delta_T^{\rm eff}
=
\delta_T
-
\frac{\beta_{TA}\beta_{AT}}{\Omega_A}.
\label{eq:deltaT_eff_definition}
\end{equation}
Equivalently, one may write \(C_h=M_{\rm eff}\gamma_{\rm eff}\), with
\begin{equation}
\gamma_{\rm eff}
=
\gamma
-
\frac{\beta_{hA}\beta_{Ah}}{M_{\rm eff}\Omega_A}.
\label{eq:gamma_eff_definition}
\end{equation}

In equation form, the reduced dynamics reads
\begin{align}
\partial_t h
&=
\left(\alpha_h q^2-C_h q^4\right)h
+
i\beta_{hT}q\,T
+\cdots ,
\label{eq:h_reduced_equation}
\\
\partial_t T
&=
i\beta_{Th}q\,h
+
\left(-\Omega_T+\delta_T^{\rm eff}q^2\right)T
+\cdots .
\label{eq:T_reduced_equation}
\end{align}
The parameter \(\Omega_T\) is the transverse gap defined in
Sec.~III.C; it vanishes in the rotationally invariant limit.

Equation~\eqref{eq:reduced_hT_matrix} is the minimal coupled
long-wavelength matrix. The coefficient \(\alpha_h\) remains the direct active
height-sector contribution and is not renormalized by elimination of the
amplitude mode at order \(q^2\). However, the leading height dynamics is
now coupled to the transverse orientational sector through the
symmetry-allowed \(iq\,\beta_{hT}\) and \(iq\,\beta_{Th}\) terms. Any
further reduction to a height-only dispersion therefore requires an
additional assumption about the transverse mode.

\subsection{Coefficients of the reduced $(h,T)$ operator}
The coefficients entering the reduced operator
\({\cal M}^{(hT)}\) of Sec.~III.F are obtained by projecting the
linearized interfacial dynamics onto the vector-valued shape functions
\(\boldsymbol{\Psi}_h\) and \(\boldsymbol{\Psi}_T\), with the
normalizations defined by the dynamical inner product introduced in
Appendix~\ref{app:projection}.

Unlike in the homogeneous bulk problem, eliminating the screened Brinkman
flow does not produce a simple local matrix correction of the form
appearing in Eq.~\eqref{eq:Lact-bulk}. The base profiles
\(\phi_0(y)\) and \(S_0(y)\) vary across the interface, and the
hydrodynamic feedback therefore appears as a nonlocal kernel in the
transverse coordinate \(y\). For the long-wavelength interfacial theory,
however, only the small-\(q\) expansion of this kernel is required.

The corresponding projected coefficients define the active contributions
to the reduced interfacial operator. The most important ones are the
direct height-sector coefficient \(\alpha_h\), the height--director
couplings \(\beta_{hT}\) and \(\beta_{Th}\), and the active contribution
to the transverse-sector coefficient \(\delta_T\). Their long-wavelength
forms are collected below, while the projection details are given in
Appendix~\ref{app:projected_matrix}.

The passive height mobility and capillary coefficient are fixed by the
equilibrium interfacial profiles. With the normalization
\begin{equation}
\Delta\phi=\phi_N-\phi_I ,
\label{eq:Delta_phi_coefficients}
\end{equation}
the effective height mobility is
\begin{equation}
M_{\rm eff}
=
M_\phi
\frac{
\displaystyle\int_{-\infty}^{\infty}dy\,\left(\phi_0'\right)^2
}{
(\Delta\phi)^2
},
\label{eq:Meff_coefficients}
\end{equation}
while the passive interfacial tension is
\begin{equation}
\gamma
=
\int_{-\infty}^{\infty}dy\,
\left[
\kappa\left(\phi_0'\right)^2
+
\frac{L}{2}\left(S_0'\right)^2
\right].
\label{eq:gamma_coefficients}
\end{equation}

The amplitude-mode correction to the capillary coefficient has been included 
in the reduced coefficient \(C_h\), defined in Eq.~\eqref{eq:Ch_definition}.

The direct active height coefficient is obtained by projecting the
hydrodynamically eliminated active operator onto the translational
height coordinate. In the local analytic expansion it may be written as
\begin{equation}
\alpha_h
=
\frac{1}{(\Delta\phi)^2}
\int dy\int dy'\,
\phi_0'(y)\,
{\cal A}_{hh}^{(2)}(y,y')\,
\phi_0'(y') .
\label{eq:alpha_h_coefficients}
\end{equation}
Here \({\cal A}_{hh}^{(2)}\) denotes the \(O(q^2)\) coefficient of the kernel of the active
hydrodynamic contribution in the height--height channel. In the aligned
geometry and with the sign convention used here, this coefficient has
the structure
\begin{equation}
\alpha_h
\sim
\zeta\,(S_N+\xi)\,I_h,
\qquad
I_h>0,
\label{eq:alpha_h_sign_structure}
\end{equation}
so that the bare active height contribution is destabilising for
extensile, flow-aligning active nematics. The value of \(I_h\) is not
universal: it depends on the interfacial profiles, the screened
hydrodynamic Green function, the friction, the viscosity, and the
boundary geometry. This coefficient is a local diffuse-interface
projection coefficient. It is distinct from the non-analytic outer-flow
contributions discussed in Sec.~IV and from height coefficients used in sharp-interface descriptions.

The leading active height--tilt coupling is defined by
\begin{equation}
{\cal M}_{hT}^{(hT)}(q)
=
i\beta_{hT}q+\cdots .
\label{eq:MhT_beta_def}
\end{equation}
Equivalently,
\begin{equation}
\beta_{hT}
=
\frac{1}{\Delta\phi}
\int dy\int dy'\,
\phi_0'(y)\,
{\cal A}_{hT}^{(1)}(y,y')\,
\psi_T(y') .
\label{eq:beta_hT_coefficients}
\end{equation}
The kernel \({\cal A}_{hT}^{(1)}\) is the \(O(q)\) coefficient of the active advective
kernel that maps the scalar transverse orientational perturbation
\(\psi_T\) onto the normal velocity of the interface. Its odd parity
follows from reflection symmetry: \(h\) is even under \(x\mapsto -x\),
whereas the transverse orientational coordinate \(T\) is odd.

The reverse coupling is defined independently by
\begin{equation}
{\cal M}_{Th}^{(hT)}(q)
=
i\beta_{Th}q+\cdots ,
\label{eq:MTh_beta_def}
\end{equation}
with
\begin{equation}
\beta_{Th}
=
\frac{1}{N_T}
\int dy\int dy'\,
\psi_T(y)\,
{\cal A}_{Th}^{(1)}(y,y')\,
\boldsymbol{\Psi}_h(y') ,
\label{eq:beta_Th_coefficients}
\end{equation}
where
\begin{equation}
N_T
=
\left\langle
\boldsymbol{\Psi}_T
\middle|
\boldsymbol{\Psi}_T
\right\rangle .
\label{eq:NT_coefficients}
\end{equation}
Here \({\cal A}_{Th}^{(1)}\) is understood as a row-vector kernel in
order-parameter space, mapping the translational perturbation
\(\boldsymbol{\Psi}_h\) into the transverse orientational equation. 

The two coefficients \(\beta_{hT}\) and \(\beta_{Th}\) are not assumed
to be equal. They are off-diagonal projections of the same linear operator, 
obtained after eliminating the screened flow field. Because activity and flow alignment make this operator non-self-adjoint, the two coefficients need not be equal.

The height projection is fixed by the conserved
composition profile, whereas the transverse projection uses the metric
of the orientational sector. The appearance of height--director
couplings is consistent with earlier passive nematic--isotropic
surface-wave theories and with recent active surface-wave models,
although here the coefficients are defined as projections of the
diffuse-interface operator.

Finally, the transverse diagonal entry is written as
\begin{equation}
{\cal M}_{TT}^{(hT)}(q)
=
-\Omega_T+\delta_T^{\rm eff}q^2+\cdots ,
\label{eq:MTT_coefficients}
\end{equation}

where $\delta_T^{\rm eff}$ is given by Eq.~\eqref{eq:deltaT_eff_definition}.
The gap \(\Omega_T\) was defined in Sec.~III.C. It vanishes in the
rotationally invariant two-dimensional limit and is finite when
orientational symmetry is broken by anchoring, elastic anisotropy,
finite size, or other explicit symmetry-breaking effects.

\subsection{Scalar and coupled interfacial regimes}

The reduced operator in Eq.~\eqref{eq:reduced_hT_matrix}
separates two issues. The coefficient \(\alpha_h\) measures the
direct local active contribution to the height sector. Whether this
coefficient controls a scalar height instability depends on the status
of the transverse orientational mode.

If the transverse mode is effectively damped, $\Omega_T>0$, it can be
eliminated at sufficiently long wavelengths. The height-like eigenvalue is then
obtained by taking the Schur complement of the transverse sector,
\begin{equation}
\mathcal{M}^{\rm eff}_{hh}(q)
=
\mathcal{M}^{(hT)}_{hh}(q)
-
\frac{
\mathcal{M}^{(hT)}_{hT}(q)
\mathcal{M}^{(hT)}_{Th}(q)
}{
\mathcal{M}^{(hT)}_{TT}(q)
}.
\label{eq:Schur_T_elimination}
\end{equation}
Using Eq.~\eqref{eq:reduced_hT_matrix}, this gives
\begin{equation}
\sigma_h(q)
=
\alpha_{\rm eff}q^2
-
C_{\rm eff}q^4
+\cdots ,
\label{eq:scalar_height_dispersion}
\end{equation}
with
\begin{equation}
\alpha_{\rm eff}
=
\alpha_h
-
\frac{\beta_{hT}\beta_{Th}}{\Omega_T}.
\label{eq:alpha_eff_definition}
\end{equation}
Thus the direct active coefficient $\alpha_h$ is not, in general, the full
coefficient controlling the scalar height-like growth rate. It is the bare
local active height-channel contribution, while $\alpha_{\rm eff}$ is the
coefficient after transverse orientational feedback has been included.

The coefficient $C_{\rm eff}$ contains the capillary stiffness $C_h$ together
with finite corrections generated by eliminating the transverse mode. To the
order displayed in Eq.~\eqref{eq:reduced_hT_matrix},
\begin{equation}
C_{\rm eff}
=
C_h
+
\frac{\beta_{hT}\beta_{Th}\delta_T^{\rm eff}}{\Omega_T^2}
+\cdots ,
\label{eq:Ceff_definition}
\end{equation}
where the ellipsis denotes additional $q^4$ contributions from higher-order
terms in the off-diagonal matrix elements.

When $C_{\rm eff}>0$, the scalar height-like mode is unstable at long
wavelengths if
\begin{equation}
\alpha_{\rm eff}>0 .
\label{eq:scalar_instability_condition}
\end{equation}
The unstable band and fastest-growing mode are then, to this order,
\begin{equation}
0<q^2<\frac{\alpha_{\rm eff}}{C_{\rm eff}},
\qquad
q_*^2=\frac{\alpha_{\rm eff}}{2C_{\rm eff}},
\qquad
\sigma_{\rm max}
=
\frac{\alpha_{\rm eff}^2}{4C_{\rm eff}} .
\label{eq:scalar_unstable_band}
\end{equation}
For $\alpha_{\rm eff}<0$, the scalar height-like mode is stable at sufficiently
long wavelengths.

In the symmetry-preserving limit discussed in Sec.~III.C,
\(\Omega_T=0\), and the scalar reduction is no longer controlled. The leading operator is then
\begin{equation}
\mathcal{M}^{(hT)}(q)
=
\begin{pmatrix}
\alpha_h q^2 & i\beta_{hT}q \\
i\beta_{Th}q & \delta_T^{\rm eff}q^2
\end{pmatrix}
+\cdots .
\label{eq:soft_T_operator}
\end{equation}
The corresponding eigenvalues are
\begin{align}
&\sigma_\pm(q)
= \nonumber \\ &
\frac{\alpha_h+\delta_T^{\rm eff}}{2}q^2
\pm
\frac{1}{2}
\left[
-4\beta_{hT}\beta_{Th}q^2
+
\left(\alpha_h-\delta_T^{\rm eff}\right)^2q^4
\right]^{1/2}
+\cdots .
\label{eq:soft_T_eigenvalues}
\end{align}
Thus, in the soft transverse regime, the leading dynamics is a genuinely
coupled height--director problem. The onset of instability cannot be assigned
to the scalar height coefficient alone.

The character of the leading pair is controlled by the sign of
$\beta_{hT}\beta_{Th}$. If
\begin{equation}
\beta_{hT}\beta_{Th}<0 ,
\end{equation}
the square root is real and the coupled sector contains an $O(|q|)$
growth--decay pair,
\begin{equation}
\sigma_\pm(q)
=
\pm
\sqrt{-\beta_{hT}\beta_{Th}}\,|q|
+
\frac{\alpha_h+\delta_T^{\rm eff}}{2}q^2
+\cdots .
\label{eq:soft_T_real_pair}
\end{equation}
If
\begin{equation}
\beta_{hT}\beta_{Th}>0 ,
\end{equation}
the leading pair is propagating,
\begin{equation}
\sigma_\pm(q)
=
\frac{\alpha_h+\delta_T^{\rm eff}}{2}q^2
\pm
i\sqrt{\beta_{hT}\beta_{Th}}\,|q|
+\cdots .
\label{eq:soft_T_propagating_pair}
\end{equation}
In this case the leading real part is controlled by the trace of the coupled operator, rather than by $\alpha_h$ alone.
The appearance of $|q|$ in Eqs.~\eqref{eq:soft_T_real_pair} and
\eqref{eq:soft_T_propagating_pair} has a different origin from the singular
hydrodynamic terms discussed below. It is a consequence of
diagonalising a local analytic matrix with a soft orientational mode,
not of a nonlocal hydrodynamic kernel.

\section{Singular hydrodynamic channel}
\label{sec:singular_hydro_channel}

The diffuse-interface theory derived in Sec.~III describes a screened
local transport regime. After hydrodynamic elimination, the projected
interfacial operator is analytic in the interfacial wavenumber \(q\), and
its coefficients are determined by the interfacial profiles, the screened
hydrodynamic Green function, and the projection onto the slow modes.

In geometries in which momentum is transported over distances large
compared with the interfacial width, or effectively unscreened over the
wavelengths of interest, the local analytic expansion is replaced,
or supplemented in a composite reduced description, by non-analytic
outer-flow contributions. These terms cannot be absorbed into the local 
diffuse-interface coefficients: they reflect the outer hydrodynamic 
boundary-value problem rather than the internal structure of the diffuse 
interface. The additive form used below is therefore a minimal modelling
decomposition, useful to classify the possible long-wavelength structures.
It is not meant to replace a full matched calculation of the crossover 
kernel when the local and singular terms originate from the same hydrodynamic response.

We write the reduced interfacial operator schematically as
\begin{equation}
\mathcal M(q)
=
\mathcal M_{\rm reg}(q)+\mathcal M_{\rm sing}(q),
\label{eq:M_reg_plus_M_sing}
\end{equation}
where \(\mathcal M_{\rm reg}\) is the analytic operator derived in
Sec.~III, and \(\mathcal M_{\rm sing}\) contains the non-analytic
contribution generated by long-ranged outer-flow transport.

A full sharp-interface theory would require solving the coupled
outer-flow, orientational, and interfacial stress-balance problem, and
could in principle generate additional (singular) height--tilt and
tilt--height couplings. Below we adopt a more restricted modelling choice:
we retain only the standard singular contribution to the geometric height
sector. This isolates the normal-velocity channel associated with the
outer flow and allows a direct comparison with the local
diffuse-interface mechanism. 

\subsection{Singular height-sector correction}

Within this minimal extension, the singular operator is taken to act only
on the height diagonal of the reduced \((h,T)\) description,
\begin{equation}
\mathcal M_{\rm sing}(q)
=
\begin{pmatrix}
\Sigma_{hh}(q) & 0\\
0 & 0
\end{pmatrix}.
\label{eq:M_sing_height_only}
\end{equation}
This form is not imposed by symmetry. Reflection symmetry would also
allow singular off-diagonal terms with the same odd parity as the local
height--tilt couplings. Equation~\eqref{eq:M_sing_height_only} should
therefore be read as a controlled truncation: it retains the leading
outer-flow contribution to the normal interfacial velocity and leaves a
full active-nematic sharp-interface projection to future work.

In a weakly screened channel-like geometry, including
Hele--Shaw/Saffman--Taylor settings, the long-wavelength height kernel
has the form
\begin{equation}
\Sigma_{hh}(q)=A_h |q|-B_h |q|q^2+\cdots .
\label{eq:singular_height_kernel}
\end{equation}
The first term is the leading outer-flow contribution and may be
stabilising or destabilising depending on imposed drive, viscosity
contrast, active-stress imbalance, or spontaneous active flow.  The
second term is the leading singular capillary regularisation.  The same
\(|q|q^2\) capillary structure appears in sharp-interface
Saffman--Taylor theory and in recent active liquid-crystal/passive-fluid
height descriptions~\cite{saffman1958,homsy1987,adkins2022,gulati2024}.
Both coefficients are fixed by the outer-flow boundary-value problem and
are not determined by the local diffuse-interface projection (see Appendix D).

The key distinction is therefore between the analytic coefficient
\(\alpha_h\), which is produced by the local diffuse-interface projection,
and the singular kernel \(\Sigma_{hh}\), which is controlled by
long-ranged incompressible momentum transport.

\subsection{Reduced $(h,T)$ operator with the singular channel}

Since the singular correction in Eq.~\eqref{eq:M_sing_height_only} acts
only in the height sector, it simply adds to the \(hh\) entry of the
analytic reduced operator derived in Sec.~III. The long-wavelength
\((h,T)\) operator becomes
\begin{align}
\mathcal M^{(hT)}(q) =
\begin{pmatrix}
\Sigma_{hh}(q)+\alpha_h q^2-C_h q^4
&
i\beta_{hT}q
\\
i\beta_{Th}q
&
-\Omega_T+\delta_T^{\rm eff}q^2
\end{pmatrix}
+\cdots ,
\label{eq:hT_with_singular_channel}
\end{align}
with \(\Sigma_{hh}(q)\) given by Eq.~\eqref{eq:singular_height_kernel}.

Equation~\eqref{eq:hT_with_singular_channel} is therefore an effective
comparison form.  The analytic terms are those obtained from the screened
diffuse-interface projection, whereas \(\Sigma_{hh}\) represents the leading
height-sector contribution of weakly screened or effectively unscreened
outer-flow transport.  Possible singular height--director couplings are
symmetry-allowed, but are not derived here.

This separation provides a useful classification of long-wavelength
interfacial regimes. In Sec.~V we use it to distinguish two independent
features of the instability: the transport channel, according to whether
the leading height dynamics is analytic or singular in \(q\), and the
modal structure, according to whether the transverse orientational mode is
damped and eliminable or remains soft and must be retained explicitly.

\section{Classification of transport channels and modal regimes}
\label{sec:classification}

The preceding sections identify two logically distinct aspects of active
nematic interfacial stability. The first is the transport mechanism.
Local diffuse-interface transport produces analytic contributions to the
projected operator, such as the active height-sector term
\(\alpha_h q^2\). Long-ranged hydrodynamic transport produces
non-analytic terms, such as \(A_h|q|\) and \(-B_h |q|q^2\).
The second aspect is the modal structure. The unstable branch may be
asymptotically scalar and height-like, or it may be a coupled
height--director mode involving the transverse orientational sector.

\subsection{Scalar height regime}

When the transverse orientational mode is effectively damped,
\(\Omega_T>0\), it may be eliminated at sufficiently long wavelengths.
The result is a scalar height-like dispersion of the form
\begin{equation}
\sigma_h(q)
=
A_h|q|+\alpha_{\rm eff}q^2
-B_h|q|q^2-C_{\rm eff}q^4+\cdots ,
\label{eq:scalar_dispersion_classification}
\end{equation}
Here \(\alpha_{\rm eff}\) is the local analytic \(q^2\) coefficient after
transverse orientational feedback has been included, Eq.~\eqref{eq:alpha_eff_definition}. The coefficient
\(A_h\) measures the leading singular hydrodynamic contribution,
\(B_h\) is the singular capillary regularisation, and \(C_{\rm eff}\)
is the local diffusive capillary regularisation.

Equation~\eqref{eq:scalar_dispersion_classification} gives a compact
classification of scalar long-wavelength instabilities.

If
\begin{equation}
A_h=0,\qquad \alpha_{\rm eff}>0,
\end{equation}
the leading destabilising contribution is local and analytic,
\begin{equation}
\sigma_h(q)\sim \alpha_{\rm eff}q^2 .
\end{equation}
The instability is then driven by the diffuse-interface channel. For
\(C_{\rm eff}>0\), the unstable band and fastest-growing mode are given by 
Eq.~\eqref{eq:scalar_unstable_band}.

If
\begin{equation}
A_h>0,
\end{equation}
the non-analytic hydrodynamic channel dominates the longest wavelengths,
since \(|q|\gg q^2\) as \(q\to0\). The instability is then controlled
asymptotically by outer-flow transport rather than by the local
diffuse-interface channel. Shorter wavelengths are regularised by the
singular capillary term \(-B_h|q|q^2\) and by the local diffusive
capillary term \(-C_{\rm eff}q^4\).

If \(A_h<0\), the scalar height branch is asymptotically stable at the
longest wavelengths. A positive \(\alpha_{\rm eff}\) may still generate a
finite-wavelength unstable band, but only if it overcomes the stabilising
singular contribution.

When both \(A_h\) and \(\alpha_{\rm eff}\) are nonzero, the measured
growth rate may reflect a crossover between the singular hydrodynamic
channel and the local analytic diffuse-interface channel,
\begin{equation}
A_h|q|\sim \alpha_{\rm eff}q^2 .
\end{equation}
Thus the dominant mechanism can depend on wavelength, screening length,
geometry, imposed drive, activity, and boundary conditions. Asymptotically,
however, a nonzero \(A_h\) controls the longest-wavelength scalar
stability, while \(\alpha_{\rm eff}\) controls the leading scalar
instability only when the singular channel is absent or subdominant.

\subsection{Coupled height--director regime}

The scalar classification above assumes that the transverse orientational
mode is effectively damped. This assumption fails when the transverse
sector is soft. In the symmetry-preserving limit discussed in Sec.~III, \(\Omega_T=0\).
The correct long-wavelength problem is therefore obtained directly from
Eq.~\eqref{eq:hT_with_singular_channel} setting \(\Omega_T=0\), rather than
by reducing to a scalar height equation.
The unstable branch, when present, is then generically a mixed
height--director mode. It cannot be characterized by a scalar height
coefficient alone.

This coupled regime preserves the distinction between transport mechanism
and modal structure. The diagonal term \(\Sigma_{hh}(q)\) identifies the
singular outer-flow contribution, while \(\alpha_h q^2\) identifies the
direct local diffuse-interface height channel. The off-diagonal terms
\(i\beta_{hT}q\) and \(i\beta_{Th}q\) determine how the height mode mixes
with the transverse orientational sector. Thus the instability may be
local in its transport mechanism even when the unstable eigenmode is not a
pure height fluctuation.

In particular, if the singular channel is absent,
\begin{equation}
\Sigma_{hh}(q)=0,
\end{equation}
the leading local coupled operator is given by Eq.~\eqref{eq:soft_T_operator}.

As shown in Sec.~III, diagonalising this analytic matrix may generate
leading eigenvalues proportional to \(|q|\), or a propagating pair,
depending on the sign of \(\beta_{hT}\beta_{Th}\). This \(|q|\)
dependence is a modal effect caused by the soft transverse sector. It
should not be confused with the non-analytic kernel produced by
long-ranged hydrodynamic transport.

\subsection{Relation to bulk stability and current experiments}

The classification above also separates interfacial instabilities from
bulk active-nematic instabilities. In the screened Brinkman regime, the
homogeneous active nematic has a finite bulk bend or splay threshold.
Below this threshold, the bulk transverse mode is linearly stable. The
interface can nevertheless become unstable through the local projected
operator: in the scalar regime through a positive \(\alpha_{\rm eff}\),
and in the coupled regime through the eigenvalues of the full
height--director block.

Above, or close to, the bulk threshold, the interpretation is less clean.
The ordered phase can generate spontaneous active currents, and in the
nonlinear regime active turbulence can advect and deform the interface.
In such cases an observed interfacial instability may be bulk-driven,
interfacial, or mixed. The distinction depends on which contribution
controls the leading growth rate: the singular outer-flow coefficient
\(A_h\), the local projected coefficients \(\alpha_{\rm eff}\) or
\(\alpha_h\), or the coupling to a soft transverse mode.

This distinction is relevant when comparing with experiments and
recent continuum descriptions. Active liquid-crystal/passive-fluid
interfaces can display traveling interfacial waves strongly coupled to
flows in the active phase~\cite{adkins2022}.  The corresponding minimal
height--director theories, and their sharp-interface extensions, are
therefore close in modal structure to the coupled descriptions discussed
here, but they also involve hydrodynamic transport mechanisms distinct
from the screened diffuse-interface limit.

Other experimentally motivated settings appear even more strongly tied
to bulk or outer-flow dynamics. Active Saffman--Taylor experiments involve
an imposed drive and a suspension state close to zero apparent viscosity
\cite{Ganesh2025}, while confined active droplets can exhibit interfacial
undulations accompanied by spontaneous internal flows
\cite{Sessa2026}. These examples are therefore best viewed as
bulk-flow-driven or coupled bulk--interface regimes, rather than as clean
realizations of the uniformly aligned, bulk-stable diffuse-interface
limit isolated here.

The local diffuse-interface instability identified in this work should therefore be regarded as a distinct limiting mechanism: it is expected to be most clearly observable below the bulk active-nematic threshold and in hydrodynamically screened geometries.
An instability observed below the bulk active-nematic threshold and in a
screened geometry would support this local mechanism. By contrast, a
growth rate controlled by a leading \(|q|\) contribution, or an
instability that appears only when the active phase develops large-scale
flow, points to a long-ranged hydrodynamic or bulk-flow-driven channel.
A propagating or mixed height--director branch would indicate that modal
structure, not only height transport, is essential. Quantitative tests of
the local mechanism require evaluating the projected coefficients and
comparing the resulting dispersion relation with controlled simulations
or experiments.

\subsection{Summary}

The main conclusions are as follows. First, projection of the
hydrodynamically screened active diffuse-interface dynamics generates a
direct active contribution to the conserved height dynamics,
\begin{equation}
M_{hh}(q)=\alpha_h q^2-C_hq^4+\cdots .
\end{equation}
This identifies a local diffuse-interface channel in which activity competes
with passive diffusive capillary relaxation.

Second, the gapped amplitude mode renormalizes the subleading \(q^4\)
height stiffness and the transverse-sector stiffness, but not the direct
active \(q^2\) height coefficient.  By contrast, if the transverse
orientational mode is gapped, the retained \(iq\) height--tilt couplings
renormalize the scalar height coefficient; if it is soft, no height-only
scalar reduction is asymptotically closed.

Third, weakly screened outer-flow transport produces non-analytic
height-sector terms of the Saffman--Taylor/Hele--Shaw type, which cannot be
absorbed into the local diffuse-interface coefficients.

Thus active nematic interfaces should be classified along two independent
axes: the transport channel, local analytic or singular hydrodynamic, and
the modal structure, scalar height-like or coupled height--director. This
classification also separates genuinely interfacial instabilities from
bulk-flow-driven or active-turbulent regimes.

\section*{Acknowledgements}

We acknowledge financial support from the Portuguese Foundation for Science and Technology (FCT) under the contracts no. UID/00618/2025 (DOI: 10.54499/UID/00618/2025), UID/PRR2/00618/2025 (DOI: 10.54499/UID/PRR2/00618/2025), UID/PRR/00618/2025 (DOI: 10.54499/UID/PRR/00618/2025), 2023.10412.CPCA.A2 (DOI 10.54499/2023.10412.CPCA.A2), and FCT/Mobility/1348751812/2024-25. RC acknowledges the financial support from FAPERJ – Fundação Carlos Chagas Filho de Amparo à Pesquisa do Estado do Rio de Janeiro (Processo SEI-260003/020878/2025).

\section*{Appendices}

\appendix

\section{Interfacial shape functions}
\label{app:shape_functions}

The use of interfacial shape functions and collective coordinates follows
standard diffuse-interface reductions of interfacial dynamics~\cite{Kawasaki1982,TurskiLanger1980,JasnowZia1987}. In this appendix we define the vector-valued shape functions
\(\Psi_h(y)\), \(\Psi_A(y)\), and \(\Psi_T(y)\), which depend only on the
transverse coordinate \(y\) and are used to construct the reduced
long-wavelength basis.
Lowercase symbols \(\psi_i(y)\) denote scalar
components of these vector-valued functions. The translational and amplitude modes are localized modes of the equilibrium interfacial operator. 
The transverse mode requires separate treatment because, in the
symmetry-preserving two-dimensional theory, it is the interfacial
representative of the bulk nematic Goldstone mode rather than a
localized interfacial bound state.

These functions identify the slow degrees of freedom retained in the
reduced theory and provide the basis onto which the full linear dynamical
operator is projected in Appendices~\ref{app:projection} and
\ref{app:projected_matrix}.

\subsection{Translational mode}

The equilibrium interfacial profiles \(\phi_0(y)\) and \(S_0(y)\)
satisfy the passive Euler--Lagrange equations.  The static Hessian in
the \((\phi,Q_1)\) sector is
\begin{equation}
\mathcal H_{\phi Q_1}
=
\begin{pmatrix}
f_{\phi\phi}(y)-\kappa \partial_y^2 & f_{\phi Q_1}(y) \\
f_{Q_1\phi}(y) & f_{Q_1Q_1}(y)-\dfrac{L}{2}\partial_y^2
\end{pmatrix},
\label{eq:H_phi1}
\end{equation}
with all local coefficients evaluated on the equilibrium profiles.

Differentiating the Euler--Lagrange equations with respect to \(y\)
gives
\begin{equation}
\mathcal H_{\phi Q_1}
\begin{pmatrix}
\phi_0'(y)\\
S_0'(y)
\end{pmatrix}
=0 .
\label{eq:translation_zero_mode}
\end{equation}

Thus the translational shape function
\begin{equation}
{\bf \Psi}_h(y)
= 
\begin{pmatrix}
\phi_0'(y)\\
S_0'(y)\\
0
\end{pmatrix}
\label{eq:psi_h_app}
\end{equation}
is the Goldstone mode associated with rigid translations of the
interface.

The passive dynamical operator in the \((\phi,Q_1)\) sector is obtained
by applying the kinetic operators to the Hessian,
\begin{equation}
\mathcal L_{\phi Q_1}^{(0)}(0;y)
=
\begin{pmatrix}
M_\phi \partial_y^2 & 0\\
0 & -\Gamma_Q
\end{pmatrix}
\mathcal H_{\phi Q_1}.
\label{eq:L_phi1}
\end{equation}

\subsection{Amplitude mode}

The amplitude mode is chosen as the lowest localized non-Goldstone mode
in the \((\phi,Q_1)\) sector, separated from the translational Goldstone
mode by the orthogonality condition imposed in the projected inner
product. In
terms of the static Hessian one may write schematically
\begin{equation}
\mathcal H_{\phi Q_1}
\begin{pmatrix}
\psi_A^\phi(y)\\
\psi_A^{Q_1}(y)
\end{pmatrix}
=
\lambda_A
\begin{pmatrix}
\psi_A^\phi(y)\\
\psi_A^{Q_1}(y)
\end{pmatrix},
\qquad
\lambda_A>0 .
\label{eq:amplitude_static_mode}
\end{equation}
where the  normalization is fixed by the kinetic inner product in Appendix~B. 
The corresponding relaxation rate in the projected dynamical
operator is denoted by \(\Omega_A>0\), so that
\begin{equation}
{\cal M}_{AA}(q=0)=-\Omega_A .
\label{eq:OmegaA_def_app}
\end{equation}

In the three-component notation used in the main text,
\begin{equation}
{\bf \Psi}_A(y)
=
\begin{pmatrix}
\psi_A^\phi(y)\\
\psi_A^{Q_1}(y)\\
0
\end{pmatrix}.
\end{equation}

\subsection{Transverse director mode}

Fluctuations of the transverse component \(Q_2\) are governed, at
\(q=0\), by the static one-dimensional Hessian
\begin{equation}
{\cal H}_T
=
-\frac{L}{2}\partial_y^2+f_{Q_2Q_2}(y),
\label{eq:HT_app}
\end{equation}
with the local coefficient evaluated on the equilibrium
profiles. In the nematic bulk, rotational invariance implies
\(f_{Q_2Q_2}\to0\). The transverse sector therefore contains the
Goldstone mode associated with a uniform rotation of the nematic
director. In the isotropic bulk, \(f_{Q_2Q_2}\) tends to a positive
constant, so transverse orientational fluctuations are massive on that
side of the interface.

In the symmetry-preserving two-dimensional theory with a single elastic
constant and no explicit orientational anchoring, the transverse mode is
identified with an infinitesimal rotation of the base state. Writing
\begin{equation}
Q_1=S\cos 2\theta,
\qquad
Q_2=S\sin 2\theta ,
\label{eq:Q1Q2_theta_app}
\end{equation}
a small rotation of the aligned state \((Q_1,Q_2)=(S_0,0)\) gives
\begin{equation}
\delta Q_2 \simeq 2S_0\,\delta\theta .
\label{eq:dQ2_rotation_app}
\end{equation}
Thus, up to normalization,
\begin{equation}
\psi_T(y)\propto S_0(y),
\label{eq:psiT_shape_app}
\end{equation}
or, in three-component notation,
\begin{equation}
{\bf \Psi}_T(y)
=
\begin{pmatrix}
0\\
0\\
\psi_T(y)
\end{pmatrix}
\propto
\begin{pmatrix}
0\\
0\\
S_0(y)
\end{pmatrix}.
\label{eq:PsiT_app}
\end{equation}

This mode decays into the isotropic phase but approaches a constant in
the nematic phase. It is therefore not a localized \(L^2\) bound state in
the infinite-domain, symmetry-preserving limit. Rather, it is the
interfacial manifestation of the bulk rotational Goldstone mode.

With finite system size, anchoring, elastic anisotropy, imposed director
boundary conditions, or other orientational symmetry-breaking effects,
the transverse sector is regularized and may acquire a small static
eigenvalue. We denote the corresponding lowest transverse shape function
by the same symbol \(\psi_T\), with the understanding that in the
symmetry-preserving limit it approaches \(\psi_T(y)\propto S_0(y)\). In
terms of the static Hessian,
\begin{equation}
{\cal H}_T\psi_T=\lambda_T\psi_T,
\qquad
\lambda_T\ge 0,
\label{eq:HT_eigenvalue_app}
\end{equation}
with \(\lambda_T=0\) in the ideal symmetry-preserving limit.

The passive dynamical operator in the transverse sector is obtained by
multiplying the static Hessian by the rotational mobility with the
relaxational sign,
\begin{equation}
{\cal L}^{(0)}_T(0;y)=-\Gamma_Q{\cal H}_T .
\label{eq:LT_HT_relation_app}
\end{equation}
Thus the corresponding projected dynamical gap is
\begin{equation}
\Omega_T=\Gamma_Q\lambda_T
\label{eq:OmegaT_lambdaT_app}
\end{equation}
for the purely relaxational transverse sector, up to the normalization
used in the projected inner product. Equivalently, in the reduced
dynamical matrix we define
\begin{equation}
{\cal M}_{TT}(q=0)=-\Omega_T .
\label{eq:MTT_gap_app}
\end{equation}
Hence \(\Omega_T=0\) in the ideal symmetry-preserving limit and
\(\Omega_T>0\) when the transverse sector is effectively damped.

Hydrodynamic backflow and flow alignment generally
generate additional \(q\)-dependent contributions to the
transverse dynamics, as already seen in the bulk aligned
geometry discussed in Sec.~\ref{Bulk_linear_modes}. In the absence of explicit
orientational symmetry breaking, however, these
contributions do not by themselves generate a \(q=0\)
restoring force.

\subsection{Asymptotic behaviour and regularisation}

The three retained functions have different asymptotic character. The
translation and amplitude modes are localized interfacial modes: they
decay into both bulk phases on the relevant correlation lengths. The
transverse mode decays into the isotropic phase, but in the
symmetry-preserving limit approaches a constant in the nematic phase.

Consequently, coefficients involving \(T\) should be understood either in
a finite system or with an explicit orientational regularisation. A
rescaling of \(\psi_T\) changes the individual coefficients
\(\beta_{hT}\), \(\beta_{Th}\), and \(\delta_T^{\rm eff}\), but not the
eigenvalues of the reduced \((h,T)\) operator when the same normalization
is used consistently.

Reflection symmetry along the interface, \(x\mapsto -x\), fixes the
parity of the projected matrix elements in \(q\). This symmetry is used
in the main text and in Appendix~\ref{app:projected_matrix} to obtain the
odd height--tilt couplings \({\cal M}_{hT},{\cal M}_{Th}\propto iq\). It should be
distinguished from any special parity of the one-dimensional shape
functions under \(y\mapsto -y\), which is not a generic symmetry of a
nematic--isotropic interface.

\section{Normalisation and inner product}
\label{app:projection}

In this appendix we define the projection inner product used in the reduction to
the slow subspace spanned by the vector-valued shape functions \(\boldsymbol{\Psi}_h\),
\(\boldsymbol{\Psi}_A\), and \(\boldsymbol{\Psi}_T\). The inner product incorporates 
the different kinetic structures of the conserved composition field and the nonconserved 
nematic fields. The purpose is to fix the normalization
conventions for the projected matrix elements entering the reduced
operator in the main text. Particular attention is required for the
translational mode, because it is associated with a conserved field and
its normalization is tied to the physical height variable and to the
effective interfacial mobility \(M_{\rm eff}\).

\subsection{Dynamical inner product}

The linearised dynamics contains one conserved field, \(\phi\), and two
nonconserved fields, \(Q_1\) and \(Q_2\). The projection metric is
therefore weighted by the inverse kinetic operators of the corresponding
linearised dynamics. For two vector-valued perturbations
\begin{equation}
\boldsymbol{\Psi}
=
\begin{pmatrix}
\psi^\phi\\
\psi^{Q_1}\\
\psi^{Q_2}
\end{pmatrix},
\qquad
\boldsymbol{\Phi}
=
\begin{pmatrix}
\varphi^\phi\\
\varphi^{Q_1}\\
\varphi^{Q_2}
\end{pmatrix},
\label{eq:B_vector_fields}
\end{equation}
we introduce the regularised inverse Laplacian
\begin{equation}
G_q=\left(-\partial_y^2+q^2\right)^{-1},
\label{eq:B_Gq}
\end{equation}
with the \(q\to0\) limit understood at finite system size, or with an
equivalent long-wavelength regularisation. The projection inner product is
then
\begin{equation}
\begin{aligned}
\left\langle
\boldsymbol{\Psi}\middle|\boldsymbol{\Phi}
\right\rangle_q
=&
\int dy\,
\frac{1}{M_\phi}\,
\psi^\phi(y)\,
\left(G_q\varphi^\phi\right)(y)
\\
&+
\int dy\,
\frac{1}{\Gamma_Q}
\left[
\psi^{Q_1}(y)\varphi^{Q_1}(y)
+
\psi^{Q_2}(y)\varphi^{Q_2}(y)
\right].
\end{aligned}
\label{eq:B_inner_product}
\end{equation}
In the passive equilibrium limit this is the metric with respect to
which the relaxational operator is self-adjoint. It also accounts for
the conserved nature of the \(\phi\) sector.

Equivalently, for modes for which the \(q\to0\) limit is regularised,
one may write the conserved part formally as
\begin{equation}
\int dy\, \psi^\phi G_0\varphi^\phi
=
\int dy\,
\left(\partial_y^{-1}\psi^\phi\right)
\left(\partial_y^{-1}\varphi^\phi\right),
\label{eq:B_inverse_derivative}
\end{equation}
with the same finite-size or finite-\(q\) regularisation understood.
This form is useful for the translational mode, but the physical height
normalisation is fixed separately by
\begin{equation}
h(x,t)=\frac{1}{\Delta\phi}
\int dy\,\delta\phi(x,y,t).
\label{eq:B_height_definition}
\end{equation}

Let
\begin{equation}
\mathsf N_{ij}
=
\langle \boldsymbol{\Psi}_i|\boldsymbol{\Psi}_j\rangle
\label{eq:gram_matrix}
\end{equation}
be the normalization, or Gram, matrix of the retained vector-valued
modes. In general, the projected operator is
\begin{equation}
{\cal M}_{ij}(q)
=
\sum_k
(\mathsf N^{-1})_{ik}
\left\langle
\boldsymbol{\Psi}_k
\left|
{\cal L}(q)
\right|
\boldsymbol{\Psi}_j
\right\rangle .
\label{eq:projected_operator_general}
\end{equation}
In the orthogonal basis used in the main text,
\(\mathsf N_{ij}=N_i\delta_{ij}\), so this reduces to
\begin{equation}
{\cal M}_{ij}(q)
=
N_i^{-1}
\left\langle
\boldsymbol{\Psi}_i
\middle|
{\cal L}(q)
\middle|
\boldsymbol{\Psi}_j
\right\rangle,
\qquad
N_i=
\left\langle
\boldsymbol{\Psi}_i
\middle|
\boldsymbol{\Psi}_i
\right\rangle .
\label{eq:B_projected_operator_orthogonal}
\end{equation}
The amplitude and transverse modes are normalized below. The
translational mode is fixed by the physical height definition, and the
corresponding normalization enters the effective height mobility
\(M_{\rm eff}\).

\subsection{Orthogonality of the retained modes}

With the inner product defined above, the retained basis may be chosen so
that
\begin{equation}
\left\langle
\boldsymbol{\Psi}_h
\middle|
\boldsymbol{\Psi}_A
\right\rangle
=
\left\langle
\boldsymbol{\Psi}_h
\middle|
\boldsymbol{\Psi}_T
\right\rangle
=
\left\langle
\boldsymbol{\Psi}_A
\middle|
\boldsymbol{\Psi}_T
\right\rangle
=0 .
\label{eq:B_orthogonality}
\end{equation}

The orthogonality of the transverse mode to the height and amplitude
modes follows from the passive block structure at \(q=0\):
\(\boldsymbol{\Psi}_h\) and \(\boldsymbol{\Psi}_A\) have support only in
the \((\phi,Q_1)\) sector, while \(\boldsymbol{\Psi}_T\) lies in the
\(Q_2\) sector. The amplitude mode is chosen in the subspace orthogonal
to the translational Goldstone mode, equivalently by diagonalising the
passive interfacial operator after removing the translational mode.

This orthogonality is a normalization convention for the reduced basis.
It should not be confused with reflection symmetry along the interface,
\(x\mapsto -x\), which fixes the parity of the projected matrix elements
in \(q\) and is used in Appendix~\ref{app:projected_matrix}.

\subsection{Normalisation of the amplitude and transverse modes}

The amplitude mode is localized at the interface and is normalized as
\begin{equation}
\left\langle
\boldsymbol{\Psi}_A
\middle|
\boldsymbol{\Psi}_A
\right\rangle
=1 .
\label{eq:B_A_normalisation}
\end{equation}

For the transverse mode the same convention can be used when the mode is
regularized by finite system size, anchoring, elastic anisotropy, imposed
director boundary conditions, or another orientational symmetry-breaking
mechanism:
\begin{equation}
\left\langle
\boldsymbol{\Psi}_T
\middle|
\boldsymbol{\Psi}_T
\right\rangle
=1 .
\label{eq:B_T_normalisation}
\end{equation}

In the ideal symmetry-preserving infinite system, the scalar transverse
profile \(\psi_T(y)\propto S_0(y)\) approaches a constant in the nematic
bulk and is not a localized normalizable bound state. In that limit the
projection on \(T\) should be understood with an explicit outer cutoff or
regularisation. The individual coefficients involving \(T\), such as
\(\beta_{hT}\), \(\beta_{Th}\), and \(\delta_T^{\rm eff}\), depend on
the chosen normalization of the transverse amplitude. The eigenvalues of
the reduced \((h,T)\) operator, however, are invariant under a consistent
rescaling of the transverse shape function and collective coordinate.

\section{Projected analytic interfacial matrix}
\label{app:projected_matrix}

In this appendix we collect the projected matrix elements entering the
regular analytic part of the long-wavelength interfacial operator, using
the shape functions and dynamical inner product defined in
Appendices~\ref{app:shape_functions} and~\ref{app:projection}. The aim
is not to evaluate all coefficients for a specific parameter set, but to
define the coefficients appearing in the reduced \((h,T)\) theory and to
display a simple screened-kernel representation of the active
height-sector and height--tilt projections.

For perturbations with in-plane wavenumber \(q\),
\begin{align}
h(x,t)&=h_q e^{\sigma t+iqx},\qquad
A(x,t)=A_q e^{\sigma t+iqx},\nonumber\\
T(x,t)&=T_q e^{\sigma t+iqx},
\end{align}
the projected dynamics is
\begin{equation}
\sigma
\begin{pmatrix}
h_q\\
A_q\\
T_q
\end{pmatrix}
=
{\cal M}(q)
\begin{pmatrix}
h_q\\
A_q\\
T_q
\end{pmatrix}.
\label{eq:app_projected_dynamics}
\end{equation}
The matrix elements are defined using the inner product of
Appendix~\ref{app:projection}. In an orthogonal basis,
\begin{equation}
{\cal M}_{ij}(q)
=
N_i^{-1}
\left\langle
\boldsymbol{\Psi}_i
\middle|
{\cal L}(q)
\middle|
\boldsymbol{\Psi}_j
\right\rangle ,
\label{eq:C_projected_matrix}
\end{equation}
with the translational normalization fixed by the physical height
definition. Throughout this appendix, \({\cal L}(q)\) and
\({\cal M}(q)\) denote the regular analytic part of the linearized
interfacial operator and its projection; singular outer-flow
contributions are not included.

Using the reflection symmetry discussed in Sec.~III.E, the leading
analytic matrix in the \((h,A,T)\) basis is
\begin{align}
&{\cal M}(q)
= \nonumber \\ & 
\begin{pmatrix}
\alpha_h q^2-M_{\rm eff}\gamma q^4
&
\beta_{hA}q^2
&
i\beta_{hT}q
\\
\beta_{Ah}q^2
&
-\Omega_A+\delta_A q^2
&
i\beta_{AT}q
\\
i\beta_{Th}q
&
i\beta_{TA}q
&
-\Omega_T+\delta_T q^2
\end{pmatrix}
+\cdots .
\label{eq:app_Mreg_three_mode}
\end{align}
Here \(\Omega_A>0\). The transverse gap \(\Omega_T\) is zero in the
rotationally invariant limit and finite when the transverse sector is
regularized, as discussed in Appendix~\ref{app:shape_functions}.

\subsection{Height-sector coefficients}

The passive height contribution follows from the usual
Cahn--Hilliard capillary relaxation of a conserved diffuse interface.
For a long-wavelength displacement, the leading composition
perturbation is
\begin{equation}
\delta\phi(x,y,t)=-h(x,t)\phi_0'(y),
\label{eq:appC_delta_phi_height}
\end{equation}
up to the sign convention used in the definition of the height mode.
The corresponding capillary chemical-potential imbalance is proportional
to the curvature,
\begin{equation}
\mu_{\rm cap}= \gamma \,\partial_x^2 h + \cdots ,
\label{eq:appC_mu_cap}
\end{equation}
so that the conserved Cahn--Hilliard current gives
\begin{equation}
\partial_t h
=
- M_{\rm eff}\gamma \,\partial_x^4 h+\cdots .
\label{eq:appC_passive_height_realspace}
\end{equation}
Equivalently, for a Fourier mode \(h_q e^{iqx+\sigma t}\),
\begin{equation}
\sigma h_q
=
- M_{\rm eff}\gamma q^4 h_q+\cdots .
\label{eq:appC_passive_height_fourier}
\end{equation}
The effective mobility of the conserved height mode is
\begin{equation}
M_{\rm eff}
=
M_\phi
\frac{\displaystyle \int_{-\infty}^{\infty}dy\,(\phi_0')^2}
{(\Delta\phi)^2},
\label{eq:appC_Meff}
\end{equation}
and the passive interfacial tension is
\begin{equation}
\gamma
=
\int_{-\infty}^{\infty}dy\,
\left[
\kappa(\phi_0')^2+\frac{L}{2}(S_0')^2
\right].
\label{eq:appC_gamma}
\end{equation}
Thus the passive conserved height contribution is
\begin{equation}
\partial_t h_q
=
- M_{\rm eff}\gamma q^4 h_q+\cdots ,
\label{eq:appC_passive_height_result}
\end{equation}
which gives the entry
\(-M_{\rm eff}\gamma q^4\) in the \(hh\) element of the
projected analytic matrix.
This is the standard long-wavelength capillary relaxation of a
conserved diffuse interface, written here with the height normalization
used in Appendix~B; see, for example, the collective-coordinate
treatments of Kawasaki and co-workers and the reviews by Langer and
Bray~\cite{Kawasaki1982,TurskiLanger1980,Bray1994}.

The direct active height coefficient is obtained by eliminating the
velocity field generated by the active stress and projecting the resulting
normal velocity onto the conserved translational mode.  For a height
displacement,
\begin{equation}
\delta\phi_q(y)=-h_q\phi_0'(y),
\label{eq:appC_delta_phi_h}
\end{equation}
with the sign fixed by the height convention used in Appendix~A.  The
normal component of the active velocity induced by this perturbation may
be written, after solving the screened Brinkman equation, in the form
\begin{equation}
v_{y,q}^{\rm act}(y)
=
\frac{h_q}{\Delta\phi}
\int dy'\,
\mathcal A_{hh}(q;y,y')\,\phi_0'(y') .
\label{eq:appC_active_velocity_kernel}
\end{equation}
The kernel \(\mathcal A_{hh}\) is the height--height part of the
hydrodynamically eliminated active operator.  Since the local
diffuse-interface contribution is analytic in \(q\), its long-wavelength
expansion begins as
\begin{equation}
\mathcal A_{hh}(q;y,y')
=
q^2\mathcal A_{hh}^{(2)}(y,y')
+O(q^4).
\label{eq:appC_Ah_expansion}
\end{equation}
Projecting the interfacial kinematic relation onto the conserved height
mode gives
\begin{equation}
\partial_t h_q
=
\frac{1}{\Delta\phi}
\int dy\,\phi_0'(y)\,v_{y,q}^{\rm act}(y)
+\cdots .
\label{eq:appC_height_projection_active}
\end{equation}
Thus the active contribution to the height equation is
\begin{equation}
\partial_t h_q
=
\alpha_h q^2 h_q+\cdots ,
\end{equation}
with
\begin{equation}
\alpha_h
=
\frac{1}{(\Delta\phi)^2}
\int dy\int dy'\,
\phi_0'(y)\,
\mathcal A_{hh}^{(2)}(y,y')\,
\phi_0'(y') .
\label{eq:appC_alpha_h_def}
\end{equation}
This is the projected form of the bulk velocity elimination described in
Sec.~II, specialized here to the translational interfacial perturbation.

For a simple unbounded Brinkman closure, the screened Green function
gives the local kernel
\begin{equation}
{\cal A}^{(2)}_{hh}(y,y')
=
\frac{\zeta}{4\eta\kappa_0}
\,[S_0(y)+\xi]\,
e^{-\kappa_0|y-y'|},
\qquad
\kappa_0=\sqrt{\Gamma/\eta},
\label{eq:app_Ahh_kernel_simple}
\end{equation}
making explicit the sign structure
\begin{equation}
\alpha_h=\zeta(S_N+\xi)I_h,\qquad I_h>0 .
\label{eq:app_alpha_h_sign}
\end{equation}
Here \(I_h\) is the positive projection integral defined by the screened
Brinkman kernel. Thus, within this closure and sign convention for activity \(\zeta\) and flow-alignment parameter
\(\xi\), the direct active height contribution is destabilising for
extensile, flow-aligning active nematics with scalar order parameter
\(S_N>0\). In more general geometries or boundary conditions, however,
the sign and magnitude of the corresponding projection coefficient must
be evaluated from the full hydrodynamic kernel.

\subsection{Amplitude and transverse couplings}

The amplitude shape function is
\begin{equation}
\boldsymbol{\Psi}_A(y)
=
\begin{pmatrix}
\psi_A^\phi(y)\\
\psi_A^{Q_1}(y)\\
0
\end{pmatrix},
\label{eq:C_PsiA}
\end{equation}
with projected relaxation rate
\begin{equation}
{\cal M}_{AA}(0)=-\Omega_A,
\qquad
\Omega_A>0 .
\label{eq:C_OmegaA}
\end{equation}

The transverse shape function is
\begin{equation}
\boldsymbol{\Psi}_T(y)
=
\begin{pmatrix}
0\\
0\\
\psi_T(y)
\end{pmatrix},
\label{eq:C_PsiT}
\end{equation}
and reduces to
\(\boldsymbol{\Psi}_T\propto (0,0,S_0)^T\) in the
symmetry-preserving limit, as described in
Appendix~\ref{app:shape_functions}.

The leading height--tilt coupling is obtained by applying the same
screened-hydrodynamic elimination used for the direct height coefficient,
but now to a transverse perturbation.  A perturbation of the transverse
mode produces an active normal velocity of the form
\begin{equation}
v_{y,q}^{(T)}(y)
=
i q\, T_q
\int dy'\,
\mathcal A_{hT}^{(1)}(y,y')\,\psi_T(y')
+O(iq^3T_q).
\label{eq:appC_vy_T_kernel}
\end{equation}
The factor \(iq\) follows from reflection symmetry: the transverse mode is
odd under reversal of the in-plane wavevector, whereas the height equation
is scalar.  Projecting the resulting normal velocity onto the conserved
height mode gives
\begin{equation}
\left.\partial_t h_q\right|_T
=
\frac{1}{\Delta\phi}
\int dy\,\phi_0'(y)\,v_{y,q}^{(T)}(y)
=
i\beta_{hT}q\,T_q+\cdots .
\label{eq:appC_ht_projection_step}
\end{equation}
Thus
\begin{equation}
\mathcal M_{hT}(q)=i\beta_{hT}q+\cdots ,
\label{eq:appC_MhT}
\end{equation}
with
\begin{equation}
\beta_{hT}
=
\frac{1}{\Delta\phi}
\int dy\int dy'\,
\phi_0'(y)\,
\mathcal A_{hT}^{(1)}(y,y')\,
\psi_T(y'),
\label{eq:appC_beta_hT}
\end{equation}
where \({\cal A}^{(1)}_{hT}\) is the \(O(q)\) coefficient of the active
advective kernel mapping a transverse orientational perturbation onto
the normal interfacial velocity.

For the same unbounded Brinkman closure used above, one finds
\begin{equation}
{\cal A}^{(1)}_{hT}(y,y')
=
\frac{\zeta}{4\eta\kappa_0}
\,e^{-\kappa_0|y-y'|}.
\label{eq:app_AhT_kernel_simple}
\end{equation}
The derivation
assumes that the transverse perturbation has a well-defined localized or
regularised \(y\)-profile. In the ideal symmetry-preserving infinite
system \(\psi_T\propto S_0(y)\) approaches a constant in the nematic
phase, and the \(q\to0\) and infinite-domain limits must be taken with
the same regularisation used to define \(N_T\).

The reverse coupling is defined independently.  The translated
interfacial perturbation \(\Psi_h\) generates, after velocity elimination
and projection onto the transverse sector, a contribution
\begin{align}
& \left.\partial_t T_q\right|_h
= \nonumber \\ &
\frac{1}{N_T}
\int dy\,\psi_T(y)\,
\left[
i q
\int dy'\,
\mathcal A_{Th}^{(1)}(y,y')\,\Psi_h(y')
\right]h_q
+\cdots .
\label{eq:appC_Th_projection_step}
\end{align}
Therefore
\begin{equation}
\mathcal M_{Th}(q)=i\beta_{Th}q+\cdots ,
\label{eq:appC_MTh}
\end{equation}
with
\begin{equation}
\beta_{Th}
=
\frac{1}{N_T}
\int dy\int dy'\,
\psi_T(y)\,
\mathcal A_{Th}^{(1)}(y,y')\,{\bf\Psi}_h(y') .
\label{eq:appC_beta_Th}
\end{equation}
Here \(\mathcal A_{Th}^{(1)}\) denotes the row-vector kernel that maps the
translated interfacial perturbation into the transverse equation.
Equivalently, if only the composition component of the height mode is
written explicitly, the corresponding factor is
\(\Psi_h^\phi=\pm\phi_0'/\Delta\phi\), with the sign fixed by the height
convention of Appendix~A.

The two coefficients \(\beta_{hT}\) and \(\beta_{Th}\) are not assumed
to be equal. The eliminated-flow operator is not self-adjoint, and the
two projections use different left weights: the height projection is
fixed by the conserved composition profile, whereas the transverse
projection uses the orientational-sector metric.

For the same unbounded Brinkman closure one may write
\begin{equation}
{\cal A}^{(1)}_{Th}(y,y')
=
\frac{\zeta}{4\eta\kappa_0}
\,{\cal P}_{Th}(y,y')\,
e^{-\kappa_0|y-y'|},
\label{eq:app_ATh_kernel_simple}
\end{equation}
where \({\cal P}_{Th}\) denotes the local profile-dependent factor
generated by the Beris--Edwards flow-alignment and projection onto the
\(Q_2\) sector. This form displays the same screened hydrodynamic range
as Eq.~\eqref{eq:app_AhT_kernel_simple}, without assuming reciprocity
between the two off-diagonal projections.

The amplitude--tilt couplings are also odd in \(q\).  Their leading
analytic contributions are defined by
\begin{equation}
\mathcal M_{AT}(q)=i\beta_{AT}q+\cdots,
\qquad
\mathcal M_{TA}(q)=i\beta_{TA}q+\cdots ,
\label{eq:appC_AT_TA_leading}
\end{equation}
where
\begin{align}
& i\beta_{AT}q
=
\frac{1}{N_A}
\left(\Psi_A\middle|\mathcal L(q)\middle|\Psi_T\right)
+\cdots ,
\nonumber \\ &
i\beta_{TA}q
=
\frac{1}{N_T}
\left(\Psi_T\middle|\mathcal L(q)\middle|\Psi_A\right)
+\cdots .
\label{eq:appC_beta_AT_TA}
\end{align}
No reciprocity between the two off-diagonal projections is assumed.
Since the amplitude mode is gapped, these coefficients enter the reduced
\((h,T)\) theory only through Schur corrections.

\subsection{Elimination of the amplitude mode}

Since the amplitude mode remains gapped,
\begin{equation}
\mathcal M_{AA}(q)=-\Omega_A+O(q^2),
\qquad
\Omega_A>0,
\label{eq:appC_A_gap}
\end{equation}
it may be eliminated perturbatively at long wavelengths.  To leading
order, the amplitude equation gives
\begin{equation}
A_q
=
\frac{1}{\Omega_A}
\left[
\beta_{Ah}q^2 h_q
+
i\beta_{AT}q\,T_q
\right]
+\cdots ,
\label{eq:appC_A_slaving}
\end{equation}
where corrections from $\sigma$ and from the $O(q^2)$ part of
$\mathcal M_{AA}$ only enter at higher order in the reduced operator.

Substituting Eq.~\eqref{eq:appC_A_slaving} into the $h$ and $T$
equations gives the Schur complement of the amplitude block.  The
resulting reduced analytic $(h,T)$ operator has the form
\begin{equation}
\mathcal M^{(hT)}(q)=
\begin{pmatrix}
\alpha_h q^2-C_h q^4
&
i\beta_{hT}q
\\
i\beta_{Th}q
&
-\Omega_T+\delta_T^{\rm eff}q^2
\end{pmatrix}
+\cdots ,
\label{eq:appC_reduced_hT}
\end{equation}
where
\begin{equation}
C_h
=
M_{\rm eff}\gamma
-
\frac{\beta_{hA}\beta_{Ah}}{\Omega_A},
\label{eq:appC_Ch}
\end{equation}
and
\begin{equation}
\delta_T^{\rm eff}
=
\delta_T
-
\frac{\beta_{TA}\beta_{AT}}{\Omega_A}.
\label{eq:appC_deltaT_eff}
\end{equation}

The amplitude mode therefore renormalizes the local $q^4$ height
stiffness and the transverse $q^2$ stiffness, but it does not
renormalize the direct active $q^2$ height coefficient.  The
height--tilt couplings generated through the amplitude mode are of
order $iq^3$ and are not retained in the leading long-wavelength
operator.

\section{Outer-flow origin of the singular height kernel}
\label{app:singular_height_kernel}

The coefficients defined in Appendix~\ref{app:projected_matrix}
determine the analytic diffuse-interface contribution to the projected
interfacial operator. In weakly screened, or effectively unscreened,
outer-flow geometries, long-ranged momentum transport generates an
additional non-analytic contribution. In the minimal extension used in
the main text, this contribution is retained only in the height sector.

\subsection{Height-sector singular structure}

We write
\begin{equation}
{\cal M}(q)
=
{\cal M}_{\rm reg}(q)
+
{\cal M}_{\rm sing}(q).
\label{eq:app_M_reg_sing}
\end{equation}
Here \({\cal M}_{\rm reg}\) is the analytic diffuse-interface operator
derived in Appendix~C, while \({\cal M}_{\rm sing}\) represents the
leading non-analytic height contribution associated with an outer-flow
transport channel. The decomposition in Eq.~\eqref{eq:app_M_reg_sing} is 
used only as an effective comparison form; it is not a controlled matched
asymptotic expansion of a single hydrodynamic regime. In the minimal
extension used here, the singular contribution is retained only in the
geometric height sector. Accordingly,
\begin{equation}
{\cal M}_{\rm sing}(q)
=
\begin{pmatrix}
\Sigma_{hh}(q) & 0 & 0\\
0 & 0 & 0\\
0 & 0 & 0
\end{pmatrix}.
\label{eq:app_Msing_three_mode}
\end{equation}
After elimination of the gapped amplitude mode, this simply adds
\(\Sigma_{hh}\) to the \(hh\) entry of the reduced \((h,T)\) operator.

The long-wavelength form retained in the main text is
\begin{equation}
\Sigma_{hh}(q)
=
A_h|q|-B_h|q|q^2+\cdots .
\label{eq:app_Sigma_hh}
\end{equation}
The coefficients \(A_h\) and \(B_h\) are not local diffuse-interface
projection coefficients. They are determined by the outer hydrodynamic
boundary-value problem, including geometry, screening, imposed drive,
viscosity contrast, active stress imbalance, and hydrodynamic boundary
conditions.

\subsection{Sketch from an outer-flow mobility}

The origin of Eq.~\eqref{eq:app_Sigma_hh} can be seen from the
standard sharp-interface argument~\cite{saffman1958,homsy1987}. Eliminating the outer pressure and
velocity fields gives a nonlocal relation between the normal interfacial
velocity and the force or pressure-jump perturbation acting on the
interface. In Fourier space this relation has the form
\begin{equation}
v_n(q)
=
\mu_{\rm out}(q)\,\delta f(q),
\label{eq:app_vn_muout}
\end{equation}
where the outer-flow mobility behaves, in a weakly screened
Hele--Shaw/Saffman--Taylor-type geometry, as~\cite{saffman1958,homsy1987}
\begin{equation}
\mu_{\rm out}(q)\sim |q|.
\label{eq:app_muout_absq}
\end{equation}
This non-analytic mobility is the signature of long-ranged
incompressible transport outside the interface.

The interfacial force perturbation contains a driving contribution and a
capillary restoring contribution,
\begin{equation}
\delta f(q)=a_1 h_q-a_3 q^2 h_q+\cdots ,
\end{equation}
where \(a_1\) represents the leading drive, viscosity-contrast,
active-stress, or spontaneous-flow imbalance, while \(a_3>0\) represents
capillary restoring forces. Combining this with
\(\mu_{\rm out}(q)\sim |q|\) gives
\begin{equation}
\partial_t h_q =
\left[A_h |q|-B_h |q|q^2+\cdots\right]h_q .
\end{equation}
which is the singular height-sector contribution used in the reduced
operator.
The same singular capillary structure, proportional to
\(|q|q^2\), appears in recent sharp-interface descriptions of active
liquid-crystal/passive-fluid interfaces~\cite{adkins2022,gulati2024}.
Here it is retained only as an outer-flow height-sector correction and is
not part of the local diffuse-interface projection.

In the present theory \(A_h\) and \(B_h\) are retained
phenomenologically. They summarize the leading effect of the outer-flow
problem and supplement, rather than renormalize, the analytic coefficients
derived from the diffuse-interface projection.

\bibliography{refs}
\bibliographystyle{apsrev4-2}

\end{document}